\documentclass[preprint,nofootinbib,longbibliography,
amsmath,amssymb,aps,prd]{revtex4-2}
\usepackage{threeparttable}
\usepackage{booktabs}
\usepackage[normalem]{ulem}

\usepackage[utf8]{inputenc}
\usepackage{amsmath}
\usepackage{amsfonts,dsfont}
\usepackage{mathrsfs}
\usepackage{cancel} 
\usepackage{amssymb,ulem}
\usepackage{romannum}
\usepackage{subfigure}
\usepackage{adjustbox}
\usepackage{dcolumn}
\usepackage{graphicx}
\usepackage{bm}
\usepackage{enumerate}
\usepackage[colorlinks=true,linkcolor=blue,urlcolor=blue,filecolor=black,
citecolor=red,pdfstartview=FitV,pdftitle={},pdfsubject={},pdfkeywords={},
pdfpagemode=None,bookmarksopen=true]{hyperref}

\usepackage{geometry}
\usepackage{booktabs}

\usepackage{xcolor}
\usepackage{float}
\usepackage{enumitem}  
\usepackage{bbold}
\usepackage{tikz}
\usepackage{multirow}
\usepackage{longtable}
\usepackage{supertabular,rotating,adjustbox}
\newcommand{\la}{\langle}
\newcommand{\ra}{\rangle}
\newcommand{\B}{{\cal B}}

\newcommand{\ov}{\overline }

\newcommand{\be}{\begin{eqnarray}}
\newcommand{\en}{\end{eqnarray}}
\newcommand{\non}{\nonumber}

\newcommand{\m}{\mathcal}

\begin{document}
	\preprint{APS/PRD}
\title{\boldmath
Topological Diagram Analysis of Charmed Baryon Decays with Vector Mesons
 }
	
\author{Yixuan Wu}
\author{Fanrong Xu}
	\email{fanrongxu@jnu.edu.cn}
\affiliation{Department of Physics, 
                College of Physics and Optoelectronic Engineering,
                Jinan University,
		Guangzhou 510632, P.R. China}
\author{Hai-Yang Cheng}
\affiliation{Institute of Physics, Academia Sinica, Taipei, 
Taiwan 11529, Republic of China}

\begin{abstract}

In this work, we further develop the application of the topological diagram approach (TDA) to charmed baryon weak decays $\m{B}_c \to \m{B} V$ with a vector meson in the final state. By incorporating the Körner-Pati-Woo theorem, we show that only five independent sets of TDA parameters are required. Relations among different decay channels arising from isospin, U-spin, and V-spin symmetries are explicitly derived within the TDA framework.  
Partial wave contributions and form factors associated with different topological diagrams are extracted from global fits to the experimental data. 
It is found that the form factors $A_2$ and $B_2$ induced from the tensor interaction of the vector mreson with octet baryons
are comparable in magnitude to $A_1$ and $B_1$, implying the importance of the tensor coupling in $\m{B}_c \to \m{B} V$ decays. 
Branching fractions for all $\B_c\to \B V$ channels are predicted, and most measured modes are found to be in good agreement with current  data. 
More physical observables, including up-down asymmetries, longitudinal polarizations as well as observables in the subsequent decays are also predicted.
Our results provide a systematic framework for understanding charmed baryon weak decays with vector mesons and can be further tested with future experiments.

\end{abstract}

	\maketitle
	\clearpage
	\clearpage
	\newpage
	\pagenumbering{arabic}

\section{Introduction}
\label{sec:intro}	

In recent years, remarkable progress has been achieved in the study of charmed baryon weak decays. An increasing number of semileptonic decays, two-body and three-body hadronic decays have been experimentally measured. The measured observables include not only branching fractions, but also polarization-related quantities, such as transverse and longitudinal decay asymmetries. These experimental developments have been comprehensively summarized in a recent review~\cite{Li:2025nzx}. Such rapid progress in experiments has simultaneously provided valuable opportunities for theoretical studies of charmed baryon physics. Indeed, many theoretical investigations have emerged in recent years, some of which were reviewed earlier in Ref.~\cite{Cheng:2021qpd}.

As bound states containing (a) charm quark(s), charmed baryons present significant theoretical challenges because the strong coupling constant $\alpha_s$  at the charm scale
cannot guarantee perturbative expansion solidly. 
Meanwhile, their weak decays, induced at the electroweak scale, exhibit features distinct from ordinary strong decay processes. Among various weak decay channels of charmed baryons, two-body hadronic decays are of particular interest, since their essential features can be more readily captured than those of multibody decays. Owing to the lack of a reliable QCD-based framework for describing two-body hadronic weak decays of charmed baryons, we previously developed a phenomenological framework, namely the topological diagram approach (TDA), for the study of two-body decays with a pseudoscalar meson in the final state~\cite{Zhong:2024zme, Zhong:2024qqs, Cheng:2024lsn}.
Although the TDA is not a first-principles approach, it nevertheless exhibits remarkable predictive power. The discovery of $\Omega_{cc}$\footnote{
On June 3, 2026, the LHCb Collaboration announced the observation of the third doubly charmed baryon, $\Omega_{cc}$, at BEAUTY 2026 \cite{YHW:Beauty2026}. Notably, the decay channel used in the discovery had been explicitly proposed in a previous theoretical study employing the topological diagrams \cite{Cheng:2020wmk}.}
 provides a compelling example, suggesting that the topological diagrams encode the underlying nonperturbative dynamics.

Compared with the decays involving a pseudoscalar meson, two-body weak decays with a vector meson in the final state were relatively less explored. However, these processes contain richer phenomenological structures due to the additional polarization degrees of freedom associated with the vector meson. Moreover, they provide an important opportunity to investigate hadronic interactions at the charm scale, such as vector and tensor interactions between the vector meson and baryons. Early theoretical studies of these decays can be traced back to Refs.~\cite{Pakvasa:1990if, Cheng:1991sn}. In recent years,  efforts have also been devoted to improving our understanding of charmed baryon dynamics through these decay channels. One commonly adopted strategy is to extract theoretical parameters from available experimental data through global fits~\cite{Hsiao:2019yur, Geng:2020zgr}. The predictive power of this approach depends strongly on both the theoretical parameterization and the available experimental inputs. A more ideal approach would rely more directly on theoretical dynamics and less on experimental data. One such attempt was presented in Ref.~\cite{Jia:2024pyb}, where final-state rescattering mechanism were considered. Nevertheless, both approaches still suffer from certain limitations at the current stage.

On the experimental side, progress keeps on steadily. In 2026, the BESIII Collaboration measured the branching fraction of the singly Cabibbo-suppressed decay $\Lambda_c^+\to \Sigma^0 K^{*+}$~\cite{BESIII:2026vjj}, obtaining
\begin{equation}
\B(\Lambda_c^+\to \Sigma^0 K^{*+}) = (1.23 \pm 0.57)\times 10^{-3}.
\end{equation}
The continuous emergence of new experimental measurements further motivates more comprehensive theoretical investigations. The TDA provides a promising framework for systematically studying charmed baryon weak decays. Building upon our previous studies of decays involving pseudoscalar mesons~\cite{Zhong:2024zme, Zhong:2024qqs, Cheng:2024lsn}, in this work we extend the TDA framework to the study of $\B_c\to\B V$ decays.

This paper is organized as follows. In Section~\ref{sec:kin}, we introduce the kinematics and relevant observables for two-body decays involving a vector meson. The TDA framework is presented in Section~\ref{sec:top}, where the decay amplitudes and the relations among different channels are explicitly derived. Numerical analyses, including the fitting strategy, results, and related discussions, are given in Section~\ref{sec:num}. Finally, a summary and conclusion are presented in Section~\ref{sec:con}.

\section{Kinematics and observables}
\label{sec:kin}	

In the weak decay processes of charmed baryons, including
$\mathcal{B}_c \to \mathcal{B} \, V$ studied in this work,
both weak and strong interactions are involved. The $(V-A)$ type weak interaction gives rise to the vector-type coupling in the baryon-baryon-vector meson interaction. On the other hand, the tensor-type interaction also plays a significant role, one example being the $N\!N\!\rho$ couplings at the hadron level. To account for both types of interactions, a general set of form factors is required, consisting of six independent parameters. However, two of them, associated with the vector meson momentum $q_\mu$, vanish due to the physical condition of the vector meson, namely $q \cdot \epsilon = 0$. As a result, four independent form factors suffice to fully describe the process, leading to the following expression for the decay amplitude:
\begin{equation}
\mathcal{M}(\mathcal{B}_c \to \mathcal{B} V) 
= \bar{u}_f(p_f)\,\epsilon^{*\mu} 
\left[ A_1 \gamma_\mu \gamma_5 
+ A_2 \frac{p_{f\mu}}{m_i} \gamma_5 
+ B_1 \gamma_\mu 
+ B_2 \frac{p_{f\mu}}{m_i} \right] u_i(p_i),
\label{eq:amp}
\end{equation}
where $\mathcal{B}_c$ ($\mathcal{B}$) denotes the initial charmed baryon (final-state octet baryon) with mass $m_i$ ($m_f$), and $V$ represents a vector meson. 
In the pole model, the form factors $A_{1,2}$ ($B_{1,2}$)  can be expressed in terms of the
parity-conserving (parity-violating) baryonic matrix elements and the vector and tensor couplings $g_{\B'\B V}$ 
and $f_{\B'\B V}$, respectively, defined by
\be
\label{eq:f1,f2}
{\cal L}_{\B'\B V} &=& g_{\B'\B V}{\rm Tr}(\ov\B' \gamma_\mu  V^\mu\B)+{  f_{\B'\B V}\over m_{\B'}+m_{\B} }
{\rm Tr}(\ov\B' \sigma_{\mu\nu}\B \,\partial^\mu V^\nu).
\en
The roles of these two couplings at the hadron level were discussed in earlier studies~\cite{Cheng:1991sn}, but the tensor-type coupling was often neglected in practical calculations. Likewise, the study in~\cite{Geng:2020zgr} only considered the vector-type interaction and neglected two of the four form factors, $A_2$ and $B_2$ except for factorizable contributions.
For completeness, this work retains all four form factors to provide a comprehensive analysis that includes both vector and tensor interactions at the hadron level.

Next, we consider the observables, the most basic of which is the decay width. The decay width can be described using two different approaches: helicity amplitudes and partial-wave amplitudes. Here, we choose the partial-wave amplitude approach~\cite{Pakvasa:1990if, Wang:2017mqp, Cheng:1991sn, Cheng:2018hwl, Geng:2020zgr}, which gives the following expression for the decay width
\footnote{The original expression of the decay width given in \cite{Pakvasa:1990if} is too small by a factor of 2.}
\begin{equation}
\Gamma = \frac{p_c}{4\pi} \frac{E_f + m_f}{m_i} 
\left[ 2\left(|S|^2 + |P_2|^2\right) 
+ \frac{E_V^2}{m_V^2} \left(|S + D|^2 + |P_1|^2\right) \right],
\label{eq:width}
\end{equation}
where $p_c$ is the magnitude of the three-momentum of the final-state baryon in the rest frame of the parent baryon, 
and $E_{f(V)} = \sqrt{p_c^2 + m_{f(V)}^2}$ denote the corresponding energy of the final-state baryon (vector meson). Among the three partial-wave amplitudes, $S$ and $D$ waves are parity-violating, while the $P$ wave is parity-conserving. The $P$-wave amplitude has additional complications as it contains contributions from both spin-flip and spin-nonflip for the baryons. The detailed expressions for the partial-wave amplitudes are given in~\cite{Pakvasa:1990if} as:
\begin{equation}
\begin{split}
S &= -A_1, \\
P_1 &= -\frac{p_c}{E_V} 
\left( \frac{m_i + m_f}{E_f + m_f} B_1 + B_2 \right), \\
P_2 &= \frac{p_c}{E_f + m_f} B_1, \\
D &= -\frac{p_c^2}{E_V (E_f + m_f)} (A_1 - A_2).
\end{split}
\label{eq:pw}
\end{equation}

To compute the branching fractions, we shall use the following charmed baryon lifetimes  \cite{ParticleDataGroup:2024cfk} 
\begin{equation} 
		\tau(\Lambda^+_c)= (202.6\pm 1.0)\,{\rm fs}, \quad
		\tau(\Xi^+_c)= (453\pm 5)\,{\rm fs}, \quad
		\tau(\Xi^0_c)= (149.8\pm1.9)\,{\rm fs}.
	\end{equation}
For the lifetime of $\Xi^0_c$, we have included the most recent LHCb measurement of $\tau(\Xi^0_c)$  \cite{LHCb:2025oww} for the average.

The up-down asymmetry of the vector meson with respect to the spin of the charmed baryon is defined as
\begin{equation}
\alpha = \frac{2 E_V^2 \,\mathrm{Re}\!\left[(S + D)^* P_1\right] 
+ 4 m_V^2 \,\mathrm{Re}\!\left(S^* P_2\right)}
{2 m_V^2 \left(|S|^2 + |P_2|^2\right) 
+ E_V^2 \left(|S + D|^2 + |P_1|^2\right)},
\label{eq:alpha}
\end{equation}
and the longitudinal polarization of the final baryon decaying from its unpolarized parent baryon \footnote{
The longitudinal polarization $P_L$	
  is denoted by $\alpha$ in the PDG \cite{ParticleDataGroup:2024cfk}, as well as in the BESIII \cite{BESIII:2022udq} and Belle \cite{Belle:2021zsy} publications. We thank Cai-Ping Jia for bringing this to our attention.
}
is given by 
\begin{equation}
P_L = \frac{2 E_V^2 \,\mathrm{Re}\!\left[(S + D)^* P_1\right] 
- 4 m_V^2 \,\mathrm{Re}\!\left(S^* P_2\right)}
{2 m_V^2 \left(|S|^2 + |P_2|^2\right) 
+ E_V^2 \left(|S + D|^2 + |P_1|^2\right)}.
\label{eq:PL}
\end{equation}
The subsequent decay of the vector meson $V \to PP$ provides additional information on the underlying dynamics. Its angular distribution, characterized by $\alpha_V$, is given by~\cite{Geng:2020zgr}:
\begin{equation}
\frac{d\Gamma_V}{d\cos\theta_V} \propto 1 + \alpha_V \cos^2\theta_V, 
\end{equation}
with
\begin{equation}
\alpha_V = \frac{E_V^2\left(|S + D|^2 + |P_1|^2\right) 
- m_V^2\left(|S|^2 + |P_2|^2\right)}
{m_V^2\left(|S|^2 + |P_2|^2\right)}.
\end{equation}

The form factors in Eq.~\eqref{eq:amp} encapsulate the full dynamical content of the decay. In general, each of them receives both factorizable and nonfactorizable contributions. Although a complete dynamical calculation is challenging, it is desirable. In the following, we establish relations among different decay channels via a topological diagrammatic approach (TDA), enabling the extraction of parameters containing dynamic information from the available experimental data.

In the pole model, the vector part of the $\B'\B V$ interaction leads to \cite{Cheng:1991sn}
\be \label{vector}
A_1 &=& -\sum_{\B_{n^*}(1/2^-)}\left[ {g_{\B_f\B_{n^*} V}b_{n^*i}\over m_i-m_n}+
{b_{fn^*}g_{\B_{n^*}\B_i V}\over m_f-m_n}\right], \non \\
B_1 &=& - \sum_{\B_n}\left[ {g_{\B_f\B_{n} V}a_{ni}\over m_i-m_n}+
{a_{fn}g_{\B_n\B_i V}\over m_f-m_n}\right],  \\
A_2 &=& B_2=0, \non 
\en
where we have considered the intermediate baryon pole states $\B_n(1/2^+), \B_n^*(1/2^-)$ and $\B_n^*(1/2^+)$ with
 the baryonic matrix elements  being defined by \cite{Cheng:1991sn}
\be
&& \la \B_i|H_W|\B_j\ra = \bar u_i(a_{ij}+b_{ij}\gamma_5)u_j, \non \\
&& \la \B_i^*(1/2^-)|H_W^{\rm pv}|\B_j\ra = ib_{i^*j}\bar u_i u_j, \\
&& \la \B_i^*(1/2^+)|H_W^{\rm pc}|\B_j\ra = a_{i^*j}\bar u_i\gamma_5 u_j. \non
\en
Since $b_{ij}\ll a_{ij}$, $a_{i^*j}\ll a_{ij}$ and $b_{i^* j}$ is of the same order as $a_{ij}$, we have kept only the leading contributions in Eq. (\ref{vector}). 

Contributions due to the tensor couplings are given by \cite{Cheng:1991sn}
\be
A_1 &=& -{m_i-m_f \over 2(m_i+m_f)}\sum_{\B_{n^*}(1/2^-)}\left[ {f_{\B_f\B_{n^*} V}b_{n^*i}\over m_i-m_n}+
{b_{fn^*}f_{\B_{n^*}\B_i V}\over m_f-m_n} \right], \non \\
A_2 &=& {2m_i \over m_i-m_f}A_1,  \non \\
B_1 &=&  {1\over 2} \sum_{\B_n}\left[ {f_{\B_f\B_n V}a_{ni}\over m_i-m_n}+
{a_{fn}f_{\B_n\B_i V}\over m_f-m_n}\right],  \\
B_2 &=& -{2m_i\over m_i+m_f}B_1.  \non 
\en
Evidently, only the tensor coupling will lead to non-vanishing form factors $A_2$ and $B_2$. 
The vector and tensor couplings $g_{\B\B V}$ and $f_{\B\B V}$, respectively,  have been evaluated using QCD sum rules \cite{Aliev:2009ei}. Comparing Tables I with II in \cite{Aliev:2009ei}, it follows that the tensor coupling is generally larger than the vector one by one order of magnitude. 

\section{Topological diagrams}
\label{sec:top}	

It is well known that charm physics remains theoretically challenging, as no fully reliable QCD-based framework has yet been established. Consequently, phenomenological approaches remain the primary practical tools for investigating these processes. Among them, the topological diagram approach (TDA) provides an intuitive and systematic framework for studying weak decays of charmed hadrons.
Previous applications of TDA to charmed baryon decays into an octet baryon and a pseudoscalar meson have been developed in Refs.~\cite{Zhong:2024zme, Zhong:2024qqs, Cheng:2024lsn}, along with studies of their direct {\it CP} violation~\cite{Cheng:2025oyr}. In the following, we extend this framework to the case where a vector meson appears in the final state.

\begin{figure}[h!]
\centering
\includegraphics[scale=1]{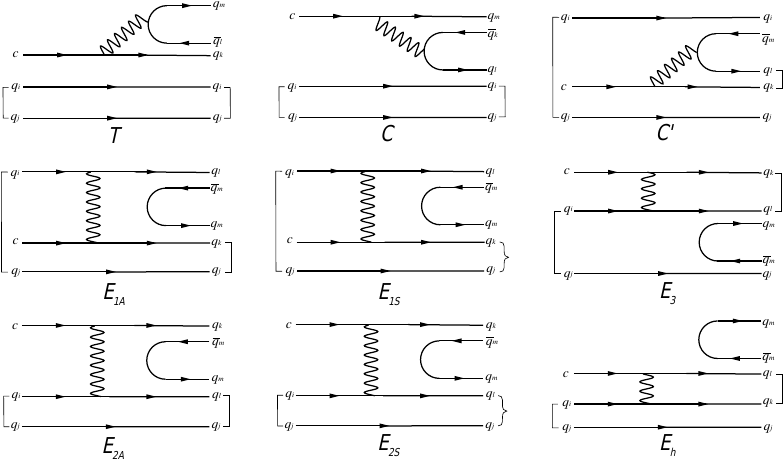}
\caption{Topological diagrams contributing to ${\cal B}_c(\bar 3)\to {\cal B}(8)V$ decays. The symbols $]$ and $\}$ denote antisymmetric and symmetric two quark states, respectively. 
}
\label{Fig:TopDiag}
\end{figure}

It is important to distinguish between topological diagrams and Feynman diagrams. In the former, only weak interactions are explicitly depicted, while the effects of strong interactions (i.e., gluon exchanges and final-state rescattering) are absorbed into the initial and final hadronic states.
In Fig.~\ref{Fig:TopDiag} we present a complete set of topological diagrams. Based on these, the general expression for the TDA amplitude can be written as
\begin{eqnarray}
\mathcal{A} &=& 
T(\mathcal{B}_c)^{ij} H_l^{km} (\mathcal{B}_8)_{ijk} V_m^l 
+ C(\mathcal{B}_c)^{ij} H_k^{ml} (\mathcal{B}_8)_{ijl} V_m^k 
+ C'(\mathcal{B}_c)^{ij} H_m^{kl} (\mathcal{B}_8)_{klj} V_i^m \nonumber \\
&& + E_{1A}(\mathcal{B}_c)^{ij} H_i^{kl} (\mathcal{B}_8)_{jkm} V_l^m 
+ E_{1S}(\mathcal{B}_c)^{ij} H_i^{kl} 
\left[(\mathcal{B}_8)_{jmk} + (\mathcal{B}_8)_{kmj}\right] V_l^m \nonumber \\
&& + E_{2A}(\mathcal{B}_c)^{ij} H_i^{kl} (\mathcal{B}_8)_{jlm} V_k^m 
+ E_{2S}(\mathcal{B}_c)^{ij} H_i^{kl} 
\left[(\mathcal{B}_8)_{jml} + (\mathcal{B}_8)_{lmj}\right] V_k^m \nonumber \\
&& + E_{3}(\mathcal{B}_c)^{ij} H_i^{kl} (\mathcal{B}_8)_{klm} V_j^m 
+ E_{h}(\mathcal{B}_c)^{ij} H_i^{kl} (\mathcal{B}_8)_{klj} V_m^m.
\label{eq:TDAamp}
\end{eqnarray}

The construction of Eq.~\eqref{eq:TDAamp} follows the same strategy as that used for $\mathcal{B}_c \to \mathcal{B} P$ 
decays~\cite{Zhong:2024zme, Zhong:2024qqs, Cheng:2024lsn}, 
with redundant degrees of freedom systematically eliminated to $9$. 
In particular, the imposed relations 
\begin{equation}
E_{1A} = -E_{2A}, \qquad
E_{1S} = -E_{2S},
\label{eq:KPW}
\end{equation}
are kept as well according to the Körner–Pati–Woo (KPW) theorem \cite{Korner:1970xq,Pati:1970fg} leading the
degrees of freedom being reduced to $7$.
The hadronic multiplets appearing in Eq.~\eqref{eq:TDAamp} are given by
\begin{equation}
\begin{split}
V_j^i &= 
\begin{pmatrix}
\frac{\rho^0 + \omega}{\sqrt{2}} & \rho^+ & K^{*+} \\ 
\rho^- & \frac{-\rho^0 + \omega}{\sqrt{2}} & K^{*0} \\ 
K^{*-} & \overline{K}^{*0} & \phi
\end{pmatrix}, \quad
(\mathcal{B}_c)^{ij} = 
\begin{pmatrix}
0 & \Lambda_c^+ & \Xi_c^+ \\ 
-\Lambda_c^+ & 0 & \Xi_c^0 \\ 
-\Xi_c^+ & -\Xi_c^0 & 0
\end{pmatrix}, \\ \\
(\mathcal{B}_8)_j^i &= 
\begin{pmatrix}
\frac{1}{\sqrt{6}} \Lambda^0 + \frac{1}{\sqrt{2}} \Sigma^0 & \Sigma^+ & p \\
\Sigma^- & \frac{1}{\sqrt{6}} \Lambda^0 - \frac{1}{\sqrt{2}} \Sigma^0 & n \\
\Xi^- & \Xi^0 & -\sqrt{\frac{2}{3}} \Lambda^0
\end{pmatrix}.
\end{split}
\end{equation}
The rank-three tensor for the baryon octet 
is defined as
$(\mathcal{B}_8)_{ijk} = \epsilon_{ijl} (\mathcal{B}_8^{T})_{\;k}^{l}$.
The relevant Cabibbo–Kobayashi–Maskawa (CKM) matrix elements are incorporated in the
tensor coefficients 
\begin{equation}
H_{2}^{31} = V_{cs}V_{ud}, \qquad 
H_{3}^{31} = V_{cs}V_{us}, \qquad 
H_{2}^{21} = V_{cd}V_{ud}, \qquad 
H_{3}^{21} = V_{us}V_{cd},
\end{equation}
which originate from the underlying weak interactions.

Among the seven TDA amplitudes $T, C, C',E_{1S}, E_{1A}, E_3$ and $E_h$ in Eq.~\eqref{eq:TDAamp}, 
there still exist two reduntant degrees of freedom through the redefinitions in terms of tilded parameters \cite{Zhong:2024qqs}:
\begin{equation}
\begin{split}
\tilde{T} &= T - E_{1S}, \qquad
\tilde{C} = C + E_{1S}, \qquad
\tilde{C'} = C' - 2E_{1S},\\
\tilde{E}_1 &= E_{1A} + E_{1S} - E_{3}, \qquad
\tilde{E}_h = E_{h} + 2E_{1S},
\end{split}
\end{equation}
so that the number of independent amplitudes is reduced to five. The resulting decay amplitudes are summarized in Tables~\ref{tab:CFamp}, \ref{tab:SCSamp}, and \ref{tab:DCS}.
This reparameterization is consistent with that employed in the $\mathcal{B}_c \to \mathcal{B} P$ case~\cite{Zhong:2024zme, Zhong:2024qqs, Cheng:2024lsn}. It should be emphasized, however, that such a redefinition is not unique. Moreover, each tilded diagrammatic parameter corresponds to four independent partial-wave amplitudes.

\begin{table}[t]\footnotesize
\centering
\caption{ The expansions of Cabbibo-favored  decay amplitudes, 
in which the common CKM factor 
$V_{cs}V_{ud}$ has been ignored. 
}.
\label{tab:CFamp}
\begin{tabular}{lll}
\hline
Channel & ~~~~~~TDA & ~~~~~$\widetilde{\mathit{\rm TDA}}$  \\
\hline

$\Lambda_c^+ \to \Xi^0 K^{*+}$&
$E_{1A}+E_{1S}-E_{3}$&
\qquad $\Tilde{E_{1}}$\\

$\Lambda_c^+ \to \Lambda \rho^+$ &  
$\frac{1}{\sqrt{6}}(-4T+C'+E_{1A}+3E_{1S}-E_{3})$&
\qquad $\frac{1}{\sqrt{6}}(-4\Tilde{T}+\Tilde{C'}+\Tilde{E_{1}})$\\

$\Lambda_c^+ \to p \bar{K}^0$&
$2C+2E_{1S}$&
\qquad $2\Tilde{C}$\\

$\Lambda_c^+ \to \Sigma^0 \rho^+$ &  
$\frac{1}{\sqrt{2}}(-C'-E_{1A}+E_{1S}+E_{3})$&
\qquad $\frac{1}{\sqrt{2}}(-\Tilde{C'}-\Tilde{E_{1}})$\\

$\Lambda_c^+ \to \Sigma^+ \rho^0$ &  
$\frac{1}{\sqrt{2}}(C'+E_{1A}-E_{1S}-E_{3})$&
\qquad $\frac{1}{\sqrt{2}}(\Tilde{C'}+\Tilde{E_{1}})$\\

$\Lambda_c^+ \to \Sigma^+ \phi$ &  
$-E_h-2E_{1S}$&
\qquad $-\Tilde{E}_h$\\

$\Lambda_c^+ \to \Sigma^+ \omega$ &  
$\frac{1}{\sqrt{2}}(-C'+E_{1A}-E_{1S}-E_{3}-2E_h)$&
\qquad $\frac{1}{\sqrt{2}}(-\Tilde{C}'+\Tilde{E_1}-2\Tilde{E}_h)$\\

$\Xi_c^0 \to \Lambda \bar{K}^{*0}$&
$\frac{1}{\sqrt{6}}(2C-C'-E_{1A}+3E_{1S}+E_{3})$&
\qquad $\frac{1}{\sqrt{6}}(2\Tilde{C}-\Tilde{C'}-\Tilde{E_{1}})$\\

$\Xi_c^0 \to \Sigma^0 \bar{K}^{*0}$&
$\frac{1}{\sqrt{2}}(2C+C'+E_{1A}+E_{1S}-E_{3})$&
\qquad $\frac{1}{\sqrt{2}}(2\Tilde{C}+\Tilde{C'}+\Tilde{E_{1}})$\\

$\Xi_c^0 \to \Sigma^+ K^{*-}$&
$-E_{1A}-E_{1S}+E_{3}$&
\qquad $-\Tilde{E_{1}}$\\

$\Xi_c^0 \to \Xi^0 \rho^0$&
$\frac{1}{\sqrt{2}}(-C'+2E_{1S})$&
\qquad $\frac{1}{\sqrt{2}}(-\Tilde{C'})$\\

$\Xi_c^0 \to \Xi^0 \phi$&
$-E_{1A}+E_{1S}+E_{3}+E_{h}$&
\qquad $-\Tilde{E}_{1}+\Tilde{E}_h$\\

$\Xi_c^0 \to \Xi^0 \omega$&
$\frac{1}{\sqrt{2}}(C'+2E_{1S}+2E_h)$&
\qquad $\frac{1}{\sqrt{2}}(\Tilde{C}'+2\Tilde{E}_h)$\\

$\Xi_c^0 \to \Xi^- \rho^+$&
$2T-2E_{1S}$&
\qquad $2\Tilde{T}$\\

$\Xi_c^+ \to \Sigma^+ \bar{K}^{*0}$&
$-2C-C'$&
\qquad $-2\Tilde{C}-\Tilde{C'}$\\

$\Xi_c^+ \to \Xi^0 \rho^+$&
$-2T+C'$&
\qquad $-2\Tilde{T}+\Tilde{C'}$\\
\hline
\end{tabular}
\end{table}

\begin{table}[tp!]\footnotesize
\centering
\caption{TDA amplitudes for singly Cabibbo-suppressed  ${\cal B}_c(\bar 3)\to {\cal B}(8)V(8+1)$ decays in which the common CKM factor 
$V_{cd}V_{ud}$ has been ignored.}
\label{tab:SCSamp}
\begin{tabular}{lll}
\hline
Channel & ~~~~~~TDA & ~~~~~$\widetilde{\mathit{\rm TDA}}$  \\
\hline
$\Lambda_c^+ \to \Lambda K^{*+}$ &  
$\frac{1}{\sqrt{6}}(-4T+C'-2E_{1A}+2E_{3})$&
\qquad $-\frac{1}{\sqrt{6}}(-4\Tilde{T}+\Tilde{C'}-2\Tilde{E_{1}})$\\

$\Lambda_c^+ \to \Sigma^0 K^{*+}$ & 
$\frac{1}{\sqrt{2}}(-C'+2E_{1S})$&
\qquad $\frac{1}{\sqrt{2}}\Tilde{C'}$\\

$\Lambda_c^+ \to \Sigma^+ K^{*0}$ &
$-C'+2E_{1S}$&
\qquad $\Tilde{C'}$\\

$\Lambda_c^+ \to p \rho^0$ &
$\frac{1}{\sqrt{2}}(-2C-C'-E_{1A}-E_{1S}+E_{3})$&
\qquad $\frac{1}{\sqrt{2}}(-2\Tilde{C}-\Tilde{C'}-\Tilde{E_{1}})$\\

$\Lambda_c^+ \to p \phi$ &  
$-2C+E_{h}$&
\qquad $-2\Tilde{C}+\Tilde{E}_{h}$\\

$\Lambda_c^+ \to p \omega$&
$\frac{1}{\sqrt{2}}(2C+C'-E_{1A}+3E_{1S}+E_{3}+2E_h)$&
\qquad $\frac{1}{\sqrt{2}}(2\Tilde{C}+\Tilde{C'}-\Tilde{E}_{1}+2\Tilde{E}_h)$\\

$\Lambda_c^+ \to n \rho^+$&
$2T-C'-E_{1A}-E_{1S}+E_{3}$&
\qquad $2\Tilde{T}-\Tilde{C'}-\Tilde{E_{1}}$\\

$\Xi_c^0 \to \Lambda \rho^0$&
$\frac{1}{2\sqrt{3}}(-2C-2C'+3E_{1S}+E_{1A}-E_{3})$&
\qquad $\frac{1}{2\sqrt{3}}(-2\Tilde{C}-2\Tilde{C'}+\Tilde{E_{1}})$\\



$\Xi_c^0 \to \Lambda \phi$&
$\frac{1}{2}(-2C+C'-2E_{1A}+2E_{3}+3E_h)$&
\qquad $\frac{1}{\sqrt{6}}(-2\Tilde{C}+\Tilde{C}'-2\Tilde{E_{1}}+3\Tilde{E}_h)$\\
%

$\Xi_c^0 \to \Lambda \omega$&
$\frac{1}{2\sqrt{3}}(2C+2C'+9E_{1S}-E_{1A}+E_{3}+3E_{h})$&
\qquad $\frac{1}{2\sqrt{3}}(2\Tilde{C}+2\Tilde{C'}-\Tilde{E_{1}}+6\Tilde{E_{h}})$\\


$\Xi_c^0 \to \Sigma^0 \rho^0$&
$\frac{1}{2}(-2C-3E_{1S}-E_{1A}+E_{3})$&
\qquad $\frac{1}{2}(-2\Tilde{C}-\Tilde{E_{1}})$\\

$\Xi_c^0 \to \Sigma^0 \phi$&
$\frac{1}{\sqrt{2}}(-2C-C'+2E_{1S}-E_{1A}+E_{3}-E_h)$&
\qquad $\frac{1}{\sqrt{2}}({-}2\Tilde{C}-\Tilde{C'}-\Tilde{E_{h}})$\\

$\Xi_c^0 \to \Sigma^0 \omega$&
$\frac{1}{2}(2C-E_{1S}-E_{1A}+E_{3}-E_{h})$&
\qquad $\frac{1}{2}(2\Tilde{C}+\Tilde{E_1}-2\Tilde{E_{h}})$\\

$\Xi_c^0 \to \Sigma^+ \rho^-$&
$E_{1A}+E_{1S}-E_{3}$&
\qquad $-\Tilde{E_1}$\\

$\Xi_c^0 \to \Sigma^- \rho^+$&
$-2T+2E_{1S}$&
\qquad $2\Tilde{T}$\\

$\Xi_c^0 \to \Xi^0 K^{*0}$&
$C'+E_{1A}-E_{1S}-E_{3}$&
\qquad $-(\Tilde{C'}+\Tilde{E_1})$\\

$\Xi_c^0 \to \Xi^- K^+$&
$2T-2E_{1S}$&
\qquad $-2\Tilde{T}$\\

$\Xi_c^0 \to p K^-$&
$-E_{1A}-E_{1S}+E_{3}$&
\qquad $\Tilde{E_1}$\\

$\Xi_c^0 \to n \bar{K}^{*0}$&
$-C'-E_{1A}+E_{1S}+E_{3}$&
\qquad $\Tilde{C'}+\Tilde{E_1}$\\

$\Xi_c^+ \to \Lambda \rho^+$&
$\frac{1}{\sqrt{6}}(2T-2C'+E_{1A}+3E_{1S}-E_{3})$&
\qquad $-\frac{1}{\sqrt{6}}(2\Tilde{T}-2\Tilde{C'}+\Tilde{E_{1}})$\\

$\Xi_c^+ \to \Sigma^0 \rho^+$&
$\frac{1}{\sqrt{2}}(-2T-E_{1A}+E_{1S}+E_{3})$&
\qquad $\frac{1}{\sqrt{2}}(2\Tilde{T}{+}\Tilde{E_{1}})$\\

$\Xi_c^+ \to \Sigma^+ \rho^0$&
$\frac{1}{\sqrt{2}}(2C-E_{1A}+E_{1S}+E_{3})$&
\qquad $\frac{1}{\sqrt{2}}(2\Tilde{C}-\Tilde{E_{1}})$\\



$\Xi_c^+ \to \Sigma^+ \phi$&
$2C+C'+2E_{1S}+E_h$&
\qquad $2\Tilde{C}+\Tilde{C'}+\Tilde{E_{h}}$\\

$\Xi_c^+ \to \Sigma^+ \omega$&
$\frac{1}{\sqrt{2}}(-2C-E_{1A}+E_{1S}+E_{3}+2E_{h})$&
\qquad $\frac{1}{\sqrt{2}}(-2\Tilde{C}-\Tilde{E_{1}}+2\Tilde{E_{h}})$\\

$\Xi_c^+ \to \Xi^0 K^{*+}$&
$-2T+C'+E_{1A}+E_{1S}-E_{3}$&
\qquad $2\Tilde{T}-\Tilde{C'}-\Tilde{E_{1}}$\\

$\Xi_c^+ \to p \bar{K}^{*0}$&
$-C'+2E_{1S}$&
\qquad $\Tilde{C'}$\\

\hline
\end{tabular}
\end{table}

\begin{table}[h]
\centering
\caption{
The expansions of doubly Cabbibo-suppressed decay amplitudes, in which the common CKM factor 
$V_{cd}V_{us}$ has been ignored.
}
\label{tab:DCS}
\begin{tabular}{lll}
\hline
Channel & ~~~~~~TDA & ~~~~~$\widetilde{\mathit{\rm TDA}}$  \\ \hline
$\Lambda_c^+ \to p K^{*0}$&
$2C+C'$&
\qquad $2\Tilde{C}+\Tilde{C'}$\\

$\Lambda_c^+ \to n K^{*+}$&
$2T-C'$&
\qquad $2\Tilde{T}-\Tilde{C'}$\\

$\Xi_c^0 \to \Lambda K^{*0}$&
$\sqrt{\frac{2}{3}}(C+C'+E_{1A}-E_{3})$&
\qquad $\sqrt{\frac{2}{3}}(\Tilde{C}+\Tilde{C'}+\Tilde{E_{1}})$\\

$\Xi_c^0 \to \Sigma^0 K^{*0}$&
$\frac{1}{\sqrt{2}}(2C+2E_{1S})$&
\qquad $\sqrt{2}\Tilde{C}$\\

$\Xi_c^0 \to \Sigma^- K^{*+}$&
$2T-2E_{1S}$&
\qquad $2\Tilde{T}$\\

$\Xi_c^0 \to p \rho^-$&
$-E_{1A}-E_{1S}+E_{3}$&
\qquad $-\Tilde{E_{1}}$\\

$\Xi_c^0 \to n \rho^0$&
$\frac{1}{\sqrt{2}}(E_{1S}+E_{1A}-E_{3})$&
\qquad $\frac{1}{\sqrt{2}}\Tilde{E_{1}}$\\

$\Xi_c^0 \to n \phi$&
$C'+E_{h}$&
\qquad $\Tilde{C}'+\Tilde{E}_{h}$\\

$\Xi_c^0 \to n \omega$&
$\frac{1}{\sqrt{2}}(2E_{1S}-E_{1A}+E_{1S}+E_{3}+2E_h)$&
\qquad $\frac{1}{\sqrt{2}}(-\Tilde{E}_{1}+2E_h)$\\

$\Xi_c^+ \to \Lambda K^{*+}$&
$\sqrt{\frac{2}{3}}(-T+C'+E_{1A}-E_{3})$&
\qquad $\sqrt{\frac{2}{3}}(-\Tilde{T}+\Tilde{C'}+\Tilde{E_{1}})$\\

$\Xi_c^+ \to \Sigma^0 K^{*+}$&
$\frac{1}{\sqrt{2}}(2T-2E_{1S})$&
\qquad $\sqrt{2}\Tilde{T}$\\

$\Xi_c^+ \to \Sigma^+ K^{*0}$&
$-2C-2E_{1S}$&
\qquad $-2\Tilde{C}$\\

$\Xi_c^+ \to p \rho^0$&
$\frac{1}{\sqrt{2}}(-E_{1A}-E_{1S}+E_{3})$&
\qquad $-\frac{1}{\sqrt{2}}\Tilde{E_{1}}$\\



$\Xi_c^+ \to p \phi$&
$C'+E_{h}$&
\qquad  $\Tilde{C}'+\Tilde{E}_{h}$\\

$\Xi_c^+ \to p \omega$&
$\frac{1}{\sqrt{2}}(-E_{1A}+3E_{1S}+E_{3}+2E_h)$&
\qquad $\frac{1}{\sqrt{2}}(-\Tilde{E_{1}}+2\tilde{E}_{h})$\\

$\Xi_c^+ \to n \rho^+$&
$-E_{1A}-E_{1S}+E_{3}$&
\qquad $-\Tilde{E_{1}}$\\
\hline
\end{tabular}
\label{tab:DCSdecay}
\end{table}

One can derive a set of model-independent relations among decay amplitudes without resorting to a detailed numerical analysis, giving
\begin{equation}
\begin{split}
& M(\Lambda_c^+ \to \Sigma^0 \rho^+) = M(\Lambda_c^+ \to \Sigma^+ \rho^0), \\
& M(\Lambda_c^+ \to \Xi^0 K^{*+}) = M(\Xi_c^0 \to \Sigma^+ K^{*-}), \\
& \sqrt{2}\, M(\Lambda_c^+ \to \Sigma^0 K^{*+}) 
= M(\Lambda_c^+ \to \Sigma^+ K^{*0})
= M(\Xi_c^+ \to p\,\overline{K}^{*0}), \\
& M(\Xi_c^0 \to \Sigma^- \rho^+) = M(\Xi_c^0 \to \Xi^- K^{*+}), \\
& M(\Xi_c^0 \to \Xi^0 K^{*0}) = M(\Xi_c^0 \to n\,\overline{K}^{*0}), \\
& \sqrt{2}\, M(\Xi_c^0 \to n \rho^0) = M(\Xi_c^0 \to p \rho^-) 
= \sqrt{2}\, M(\Xi_c^+ \to p \rho^0) = M(\Xi_c^+ \to n \rho^+), \\
& \sqrt{2}\, M(\Xi_c^0 \to \Sigma^0 K^{*0}) = M(\Xi_c^+ \to \Sigma^+ K^{*0}), \\
& \sqrt{2}\, M(\Xi_c^+ \to \Sigma^0 K^{*+}) = M(\Xi_c^0 \to \Sigma^- K^{*+}),
\end{split}
\label{eq:sym}
\end{equation}
valid up to a sign. 
Such relations which hold at the amplitude level 
can be interpreted as consequences of isospin symmetry,
 U-spin symmetry and V-spin symmetry.
%
They provide non-trivial consistency checks for both theoretical frameworks and experimental measurements.
Our next task then is to extract the TDA parameters from the current data.


\section{Numerical analysis}
\label{sec:num}	

Benefiting from recent experimental advances at BESIII, Belle~II and LHCb, the accumulated data on $\mathcal{B}_c \to \mathcal{B} V$ decays now can support a comprehensive global fit. In this section, based on the latest experimental results, we perform a numerical analysis of Cabibbo-favored (CF), singly Cabibbo-suppressed (SCS), and doubly Cabibbo-suppressed (DCS) decay modes. 
Our theoretical predictions are confronted with available experimental measurements. In addition, we will present a comparison with results obtained by other theoretical approaches
in the end.

\subsection{Experimental status}
\label{subsec:exp}	

At present, more than 20 experimental measurements are available, including branching fractions and decay asymmetries. The dataset contains both recent and earlier results. 
For example, in 2025 the LHCb Collaboration measured the decay asymmetry of $\Xi_c^+ \to p\,\overline{K}^{*0}$ to be
$\alpha(\Xi_c^+ \to p\,\overline{K}^{*0}) = 0.613 \pm 0.065$~\cite{LHCb:2025hul}.
There are also  earlier measurements dating back more than two decades. An example is the branching fraction of $\Lambda_c^+ \to \Sigma^+ K^{*0}$ which was measured by the FOCUS Collaboration in 2002 to be
$\mathcal{B}(\Lambda_c^+ \to \Sigma^+ K^{*0}) = (3.5 \pm 1.0)\times 10^{-3}$~\cite{FOCUS:2002rvb}.
All available data, summarized as Table~\ref{tab:Exp}, 
are used as inputs for the global fit.

\begin{table}[tp!]%
    \centering
    \resizebox{\textwidth}{!} 
{
    \begin{threeparttable}
		\caption{
        Experimental measurements of the branching fractions and the longitudinal  decay asymmetry $P_L$ in the decays of $\mathcal{B}_c \to \B V$.
        The order of magnitude of the branching fractions in the first, second and third rows is $10^{-2},10^{-3}$ and $10^{-4}$, respectively. The data are taken from Ref.
        \cite{ParticleDataGroup:2024cfk} unless specified. }
		\begin{tabular} { l  c c c c c c} 
        \hline
	\multirow{2}{*}{\rm Mode} & \multicolumn{5}{c}{\rm $\mathcal{B}$} 
	& \multirow{2}{*} {$P_L$} 
	\\
    \cline{2-6}
    & {\rm PDG} & {\rm BES \uppercase\expandafter{\romannumeral 3}} & {\rm Belle } & {\rm LHCb } & {\rm Average } &  \\
    \hline
    $\Lambda^+_c \to p \overline{K}^{*0}$   &$1.40\pm0.07$  &$  $ &$  $ &$  $ &$1.40\pm0.07$ 
    &
    ${{ -0.835\pm0.215}}$ 
    \cite{LHCb:2022sck,Li:2025nzx}\tnote{a}\\
    $\Lambda^+_c \to \Lambda \rho^+$   &$4.1\pm0.5$  &$ $ &$ $ &$ $ &$4.1\pm0.5$   & $-0.76\pm0.07$ \\
    $\Lambda^+_c \to \Sigma^+\rho^0$   &$<1.7$    &$ $ &$ $ &$ $ &$<1.7$  \\
    $\Lambda^+_c \to \Sigma^+\omega$   &$1.72\pm0.20$  &$ $ &$  $ &$  $ &$1.72\pm0.20$  \\
    $\Xi^+_c \to \Sigma^+ \overline{K}^{*0}$ & $2.3\pm1.1$ &$ $ &$  $ &$  $ & $2.3\pm1.1$ \\
    \hline
    $\Lambda^+_c \to p\phi$   &$1.05\pm0.14$ & $1.21\pm0.14$ \cite{BESIII:2026dkj}
     &$ $ &$ $ &$1.13\pm0.10$  \\
    $\Lambda^+_c \to \Sigma^+\phi$   &$4.0\pm0.5$   &$ $ &$ $&$ $ &$4.0\pm0.5$ \\
    $\Lambda^+_c \to \Sigma^+ K^{*0}$   &$3.5\pm1.0$  &$ $ &$ $ &$ $ &$3.5\pm1.0$ \\
    $\Lambda^+_c \to p\rho^0$   &$1.5\pm0.4$ &$ $ &$  $&$ $ &$1.5\pm0.4$  \\
    $\Lambda^+_c \to \Lambda K^{*+}$ &$ $ &$1.29\pm0.44$ \cite{BESIII:2024xny}\tnote{b}
    &$ $ &$ $ &$1.29\pm0.44$\\
     ${{\Lambda^+_c \to \Sigma^0 K^{*+}}}$ &$ $ &${{1.23\pm0.57}}$ \cite{BESIII:2026vjj} &$ $ &$ $ &${{1.23\pm0.57}}$\\
    $\Xi^+_c \to p\overline{K}^{*0}$   &$3.3\pm1.7$ &$ $ &$  $ &$  $ &$3.3\pm1.7$
    &
    ${{0.303\pm0.121}}$
    \cite{LHCb:2025hul,Li:2025nzx}\tnote{a} \\
    $\Xi^+_c \to \Sigma^+\phi$   &$<3.2$ &$ $ &$  $ &$  $ &$<3.2$ \\
    $\Xi^0_c \to \Lambda\overline{K}^{*0}$   &$2.6\pm0.6$ &$ $ &$  $ &$  $ &$2.6\pm0.6$ & $0.15\pm0.22$ \\
    $\Xi^0_c \to \Sigma^0 \overline{K}^{*0}$ & $9.9\pm1.9$  &$ $ &$  $ &$  $ & $9.9\pm1.9$ \\
    $\Xi^0_c \to \Sigma^+K^{*-}$   &$4.9\pm1.3$ &$ $ &$  $ &$  $ &$4.9\pm1.3$ & $-0.50\pm0.30$\\
    \hline
    $\Lambda^+_c \to p\omega$ & $8.9\pm1.1$ & $ $ & $ $ & $9.82\pm3.30$ \cite{LHCb:2024hju}& $9.0\pm1.0$ & $ $ \\
    $ \Xi^+_c \to p\phi$   &$1.2\pm0.6$ & &$  $ &$ $ & $1.2\pm0.6$ & $ $ \\
    $ \Xi^0_c \to \Lambda\phi$   &$4.9\pm1.3$ &$ $ &$  $ &$ $ & $4.9\pm1.3$ & $ $ \\
    \hline
	\end{tabular} 	
    \begin{tablenotes}
    \item[a]
	The numbers quoted here have been recalculated by the authors of \cite{Li:2025nzx} based on the data reported in \cite{LHCb:2022sck, LHCb:2025hul}, after reconciling the different conventions for the asymmetry parameter $\alpha$ adopted by BESIII and LHCb. In the present work, this quantity is denoted by the longitudinal polarization $P_L$.    
    \item[b]
    Among all the three results in \cite{BESIII:2024xny},  the branching fractions $2.40\pm0.59$, $5.21\pm0.75$ and $1.29\pm0.44$ correspond $\theta_0=0$ (no interference), $109^\circ$ and $221^\circ$, respectively.
    \end{tablenotes}
    \end{threeparttable}
   \label{tab:Exp}
  } 
\end{table}

\subsection{Numerical results}
\label{subsec:num}

As discussed above, there are five independent topological diagrammatic parameters, each of which corresponds to four partial-wave amplitudes. Consequently, a total of 20 real parameters are involved in the fit.
In contrast, it has been argued in Ref.~\cite{Geng:2020zgr} that two of the four partial waves are subdominant and may be neglected. Whether such a simplification is justified can ultimately be tested against experimental data.
Given that the current fit already involves 20 parameters (considering only the magnitudes of the partial-wave amplitudes), it is not feasible at present to include their relative phases. As a result, predictions for CP-violating observables cannot be addressed at this stage.

The $\chi^2$ function is defined as
\begin{equation}
\chi^2 = \sum_i \frac{\big(\mathcal{B}^{\rm TDA}_i - \mathcal{B}^{\rm exp}_i\big)^2}{\Delta_i^2}
+ \sum_i \frac{\big(\alpha^{\rm TDA}_i - \alpha^{\rm exp}_i\big)^2}{\delta_i^2},
\end{equation}
where $\mathcal{B}^{\rm TDA}$ and $\alpha^{\rm TDA}$ are calculated using Eqs.~\eqref{eq:width} and ~\eqref{eq:alpha}, respectively, with amplitudes listed in Tables~\ref{tab:CFamp}, \ref{tab:SCSamp}, and \ref{tab:DCS}. 
For the uncertainties of the branching fractions ($\Delta$) and decay asymmetries ($\delta$), we assume that the experimental errors dominate and are uncorrelated. In total, 24 observables are included in the fit with 20 free parameters.

Using the package {\texttt{iminuit}}, we obtain a global minimum
 of 
 $\chi^2_{\rm min}/{\rm d.o.f.} = 19.80/(24 - 20) \approx 4.95.$
The fitted parameters, corresponding to the five independent sets of topological amplitudes with four different partial-wave components, are listed in Table~\ref{tab:fitpara}. 
The associated covariance matrix, which characterizes correlations 
among fitted parameters, can also be obtained using {\texttt{iminuit}}, yielding
\begin{eqnarray}
&&C= \nonumber\\
&&\scalebox{0.45}{$\displaystyle
\begin{pmatrix}
0.776 & 0.01 & 0.02 & 0.02 & 0.02 & -0.1 & 0.004 & -0.04 & 0.01 & 0.02 & -0.34 & 0.00 & -0.02 & -0.00 & -0.01 & -3.2 & -0.001 & -0.0 & -0.02 & -0.01 \\
0.01 & 0.0435 & -0.03 & 0.05 & 0.01 & -0.11 & -0.009 & -0.05 & 0.04 & 0.07 & -0.03 & 0.04 & -0.04 & 0.04 & -0.00 & 0.03 & -0.030 & 0.07 & -0.08 & -0.08 \\
0.02 & -0.03 & 0.0806 & -0.02 & 0.03 & 0.08 & 0.018 & -0.04 & -0.01 & -0.02 & 0.01 & -0.04 & -0.00 & -0.05 & 0.00 & -0.07 & 0.027 & -0.20 & 0.05 & 0.07 \\
0.02 & 0.05 & -0.02 & 0.0811 & 0.04 & -0.11 & 0.001 & -0.11 & 0.05 & 0.08 & -0.04 & 0.04 & -0.07 & 0.03 & -0.02 & 0.00 & -0.024 & 0.00 & -0.10 & -0.09 \\
0.02 & 0.01 & 0.03 & 0.04 & 0.0753 & 0.02 & 0.014 & -0.09 & 0.02 & 0.04 & -0.01 & -0.00 & -0.04 & -0.04 & -0.07 & -0.05 & 0.007 & -0.13 & -0.01 & -0.06 \\
-0.1 & -0.11 & 0.08 & -0.11 & 0.02 & 0.415 & 0.037 & 0.10 & -0.08 & -0.17 & 0.13 & -0.11 & 0.10 & -0.17 & -0.04 & 0.3 & 0.085 & -0.2 & 0.22 & 0.18 \\
0.004 & -0.009 & 0.018 & 0.001 & 0.014 & 0.037 & 0.0114 & -0.019 & -0.001 & -0.015 & 0.004 & -0.011 & -0.004 & -0.025 & -0.015 & -0.024 & 0.016 & -0.056 & 0.017 & 0.015 \\
-0.04 & -0.05 & -0.04 & -0.11 & -0.09 & 0.10 & -0.019 & 0.22 & -0.08 & -0.11 & 0.05 & -0.03 & 0.12 & 0.00 & 0.04 & 0.06 & 0.012 & 0.17 & 0.11 & 0.09 \\
0.01 & 0.04 & -0.01 & 0.05 & 0.02 & -0.08 & -0.001 & -0.08 & 0.0406 & 0.06 & -0.03 & 0.03 & -0.05 & 0.02 & -0.01 & 0.01 & -0.019 & 0.01 & -0.07 & -0.07 \\
0.02 & 0.07 & -0.02 & 0.08 & 0.04 & -0.17 & -0.015 & -0.11 & 0.06 & 0.129 & -0.05 & 0.06 & -0.08 & 0.07 & -0.01 & 0.03 & -0.050 & 0.04 & -0.13 & -0.13 \\
-0.34 & -0.03 & 0.01 & -0.04 & -0.01 & 0.13 & 0.004 & 0.05 & -0.03 & -0.05 & 0.166 & -0.02 & 0.04 & -0.03 & 0.00 & 1.33 & 0.018 & -0.03 & 0.06 & 0.05 \\
0.00 & 0.04 & -0.04 & 0.04 & -0.00 & -0.11 & -0.011 & -0.03 & 0.03 & 0.06 & -0.02 & 0.0442 & -0.03 & 0.05 & -0.00 & 0.04 & -0.026 & 0.10 & -0.08 & -0.08 \\
-0.02 & -0.04 & -0.00 & -0.07 & -0.04 & 0.10 & -0.004 & 0.12 & -0.05 & -0.08 & 0.04 & -0.03 & 0.0766 & -0.03 & 0.02 & 0.01 & 0.019 & 0.03 & 0.09 & 0.08 \\
-0.00 & 0.04 & -0.05 & 0.03 & -0.04 & -0.17 & -0.025 & 0.00 & 0.02 & 0.07 & -0.03 & 0.05 & -0.03 & 0.094 & 0.04 & 0.07 & -0.044 & 0.16 & -0.09 & -0.07 \\
-0.01 & -0.00 & 0.00 & -0.02 & -0.07 & -0.04 & -0.015 & 0.04 & -0.01 & -0.01 & 0.00 & -0.00 & 0.02 & 0.04 & 0.138 & 0.03 & -0.016 & 0.04 & 0.01 & 0.13 \\
-3.2 & 0.03 & -0.07 & 0.00 & -0.05 & 0.3 & -0.024 & 0.06 & 0.01 & 0.03 & 1.33 & 0.04 & 0.01 & 0.07 & 0.03 & 13.1 & -0.033 & 0.2 & -0.05 & -0.05 \\
-0.001 & -0.030 & 0.027 & -0.024 & 0.007 & 0.085 & 0.016 & 0.012 & -0.019 & -0.050 & 0.018 & -0.026 & 0.019 & -0.044 & -0.016 & -0.033 & 0.0327 & -0.075 & 0.057 & 0.049 \\
-0.0 & 0.07 & -0.20 & 0.00 & -0.13 & -0.2 & -0.056 & 0.17 & 0.01 & 0.04 & -0.03 & 0.10 & 0.03 & 0.16 & 0.04 & 0.2 & -0.075 & 0.564 & -0.13 & -0.14 \\
-0.02 & -0.08 & 0.05 & -0.10 & -0.01 & 0.22 & 0.017 & 0.11 & -0.07 & -0.13 & 0.06 & -0.08 & 0.09 & -0.09 & 0.01 & -0.05 & 0.057 & -0.13 & 0.177 & 0.16 \\
-0.01 & -0.08 & 0.07 & -0.09 & -0.06 & 0.18 & 0.015 & 0.09 & -0.07 & -0.13 & 0.05 & -0.08 & 0.08 & -0.07 & 0.13 & -0.05 & 0.049 & -0.14 & 0.16 & 0.276
\end{pmatrix}
$}.\nonumber
\end{eqnarray}

\begin{table}[t]
\centering
\caption{The fitted partial-wave components of TDA amplitudes in units of $10^{-6}$.}
\label{tab:fitpara}
\setlength{\tabcolsep}{12pt}
\renewcommand{\arraystretch}{1.3}
\begin{tabular}{|l|rr rr|}
\hline
 & $S$~~~~ & $P_1$~~~~ & $P_2$~~~~ & $D$~~~~ \\
\hline
$\tilde{T}$   & $0.40 \pm 0.90$ & $-0.10 \pm 0.60$ & $0.60 \pm 0.40$ & $-0.00 \pm 4.00$ \\
$\tilde{C}$   & $-0.01 \pm 0.21$ & $0.29 \pm 0.11$ & $0.12 \pm 0.21$ & $-0.23 \pm 0.18$ \\
$\tilde{C'}$  & $0.54 \pm 0.28$ & $-0.00 \pm 0.50$ & $-0.14 \pm 0.28$ & $-0.20 \pm 0.80$ \\
$\tilde{E}_{1}$ & $-0.23 \pm 0.28$ & $0.46 \pm 0.20$ & $-0.07 \pm 0.31$ & $0.10 \pm 0.40$ \\
$\tilde{E}_h$ & $-0.35 \pm 0.27$ & $0.40 \pm 0.40$ & $-0.20 \pm 0.40$ & $0.70 \pm 0.50$ \\
\hline
\end{tabular}
\end{table}

Applying the relations given by Eq. (\ref{eq:pw}) between partial waves and form factors, we have also performed 
global fits to the  topological amplitudes in four different form-factor components. The results are listed in Table~\ref{tab:fitpara_2}. However, it should be stressed that Tables ~\ref{tab:fitpara} and ~\ref{tab:fitpara_2}
are not correlated with each other as they are fitted independently.  

\begin{table}[t]
\centering
\caption{The fitted form-factor components of TDA amplitudes in units of $10^{-6}$.}
\label{tab:fitpara_2}
\setlength{\tabcolsep}{12pt}
\renewcommand{\arraystretch}{1.3}
\begin{tabular}{|l|rr|rr|}
\hline
 & $A_1$~~~~ & $B_1$~~~~ & $A_2$~~~~ & $B_2$~~~~ \\
\hline
$\tilde{T}$   & $0.41 \pm 0.31$ & $-2.5 \pm 1.1$ & $3.00 \pm 11.00$ & $3.00 \pm 4.00$ \\
$\tilde{C}$   & $-0.13 \pm 0.09$ & $0.01 \pm 0.35$ & $-2.40 \pm 0.40$ & $-0.30 \pm 0.50$ \\
$\tilde{C'}$  & $0.33 \pm 0.10$ & $-1.00 \pm 0.70$ & $4.20 \pm 1.40$ & $1.00 \pm 0.90$ \\
$\tilde{E}_{1}$ & $0.09 \pm 0.12$ & $1.00 \pm 0.70$ & $-1.80 \pm 0.90$ & $-1.80 \pm 1.10$ \\
$\tilde{E}_h$ & $0.40 \pm 0.04$ & $1.70 \pm 0.70$ & $0.70 \pm 2.10$ & $-2.40 \pm 1.10$ \\
\hline
\end{tabular}
\end{table}

Several observations can be made from Table~\ref{tab:fitpara_2}:
(1) The uncertainties of the fitted parameters remain sizable due to the limited amount of available data. Significant improvement is expected with the rapid progress of ongoing experiments. In this sense, establishing a systematic fitting framework for $\mathcal{B}_c \to \mathcal{B} V$ decays is both timely and well motivated.
(2) Taking the parameters associated with $\tilde{E}_h$ as an example, the fitted value 
$B_2(\tilde{E}_h) = -2.40\pm 1.10$ 
is comparable in magnitude to 
$B_1(\tilde{E}_h) = 1.70 \pm 0.70$ 
(see Table \ref{tab:fitpara_2}). This indicates that neglecting a subset of 
form factor
contributions may lead to a loss of important dynamical effects and is therefore not justified.


\begin{table}[t]
\centering
\caption{Predictions of observables in Cabibbo-favored decays. }
\label{tab:numCF}
\setlength{\tabcolsep}{10pt}
\begin{tabular}{l rrrr}
\toprule
{\rm Mode}&{\rm $10^{2}\mathcal{B}$}~~~~ &{$\alpha$}~~~~&	{$P_L$}~~~~	&{$\alpha_V$}~~~~\\
\midrule
$	\Lambda_c^+	\to	\Xi^0	K^{*+}	$	&	$0.28\pm0.09$	&	$-0.13 ^{+0.42}_{-0.46}$	&	$-0.49\pm0.34$	&	$3.12\pm14.33$	\\
$	\Lambda_c^+	\to	\Lambda	\rho^+	$	&	$4.10\pm0.50$	&	$0.73 ^{+0.14}_{-0.54}$	&	$-0.76\pm0.07$	&	$-0.87\pm0.96$	\\
$	\Lambda_c^+	\to	p	\overline{K}^{*0}   	$	&	$1.40\pm0.07$	&	$-0.88 ^{+0.23}_{-0.07}$	&	$-0.84\pm0.21$	&	$13.16 ^{+27.45}_{-8.95}$	\\
$	\Lambda_c^+	\to	\Sigma^0	\rho^+	$	&	$0.45\pm0.14$	&	$0.10 ^{+0.42}_{-0.88}$	&	$0.89\pm0.24$	&	$1.67\pm11.37$	\\
$	\Lambda_c^+	\to	\Sigma^+	\rho^0   	$	&	$0.46\pm0.15$	&	$0.10 ^{+0.42}_{-0.88}$	&	$0.89\pm0.24$	&	$1.68\pm11.41$	\\
$	\Lambda_c^+	\to	\Sigma^+	\phi  	$	&	$0.41\pm0.05$	&	$0.81 ^{+0.14}_{-0.27}$	&	$-0.14 ^{+0.50}_{-0.43}$	&	$0.40\pm2.76$	\\
$	\Lambda_c^+	\to	\Sigma^+	\omega	$	&	$1.67\pm0.17$	&	$0.27 ^{+0.35}_{-0.40}$	&	$0.41\pm0.55$	&	$5.36\pm26.46$	\\
$	\Xi_c^0	\to	\Lambda	\overline{K}^{*0}   	$	&	$0.29\pm0.06$	&	$-0.64 ^{+0.55}_{-0.20}$	&	$0.16\pm0.22$	&	$1.77\pm6.84$	\\
$	\Xi_c^0	\to	\Sigma^0	\overline{K}^{*0}   	$	&	$1.02\pm0.17$	&	$-0.36 ^{+0.18}_{-0.53}$	&	$-0.41\pm0.53$	&	$8.35 ^{+18.41}_{-5.18}$	\\
$	\Xi_c^0	\to	\Sigma^+	K^{*-}	$	&	$0.51\pm0.13$	&	$-0.20 ^{+0.42}_{-0.42}$	&	$-0.48\pm0.30$	&	$4.76\pm20.01$	\\
$	\Xi_c^0	\to	\Xi^0	\rho^0  	$	&	$0.50\pm0.22$	&	$-0.40 ^{+0.48}_{-0.36}$	&	$0.31\pm0.14$	&	$-0.32\pm2.43$	\\
$	\Xi_c^0	\to	\Xi^0	\phi  	$	&	$0.20\pm0.10$	&	$-0.04 ^{+0.41}_{-0.43}$	&	$-0.56\pm0.66$	&	$4.22\pm28.06$	\\
$	\Xi_c^0	\to	\Xi^0	\omega	$	&	$1.83\pm0.34$	&	$0.84 ^{+0.10}_{-0.39}$	&	$0.60\pm0.53$	&	$5.09\pm18.46$	\\
$	\Xi_c^0	\to	\Xi^-	\rho^+  	$	&	$4.66\pm1.39$	&	$0.88 ^{+0.06}_{-0.55}$	&	$-0.93\pm0.96$	&	$-0.95\pm1.64$	\\
$	\Xi_c^+	\to	\Sigma^+	\overline{K}^{*0}   	$	&	$3.58\pm1.01$	&	$0.03 ^{+0.28}_{-0.79}$	&	$-0.40\pm0.47$	&	$0.69\pm4.88$	\\
$	\Xi_c^+	\to	\Xi^0	\rho^+  	$	&	$12.55\pm5.02$	&	$0.36 ^{+0.36}_{-0.34}$	&	$-0.31 ^{+0.10}_{-0.29}$	&	$-0.95\pm2.09$	\\

\hline
\end{tabular}
\end{table}

The observables, including branching fractions, up-down asymmetries, longitudinal polarizations, and polarization parameters in subsequent decays, can be directly computed using the fitted parameters. In particular, the uncertainty of the $j$-th observable $f_j$, with the estimated parameters denoted by $\theta$, is given by
\begin{equation}
\sigma_{j}^2 = \sum_{i=1}^n \sum_{k=1}^n J_{ji} C_{ik} J_{jk},
\end{equation}
where the Jacobian matrix $J_{ji} = \frac{\partial f_j}{\partial \theta_i}$ can be evaluated using the \texttt{jacobi} package. The expected hierarchical pattern of branching fractions for CF, SCS, and DCS processes is reproduced, with typical magnitudes of $\mathcal{O}(10^{-2})$, $\mathcal{O}(10^{-3})$, and $\mathcal{O}(10^{-4})$, respectively, as summarized in Tables~\ref{tab:numCF}, \ref{tab:numSCS}, and \ref{tab:numDCS}.

We will  choose a few CF and SCS processes as examples to illustrate the interpretations we can draw from these theoretical predictions. Among the sixteen CF decay modes, four of them have been experimentally measured. The yet-to-be-measured mode, $\Xi_c^+ \to \Xi^0 \rho^+$, is of particular interest due to its large predicted branching fraction, 
$\mathcal{B}(\Xi_c^+ \to \Xi^0 \rho^+) = (12.55 \pm 5.02) \times 10^{-2}$. 
 A measurement of this channel is highly anticipated in the near future.

Meanwhile, the polarization observables $\alpha$ and $P_L$ provide complementary information:
\begin{enumerate}
    \item Modes such as 
    $\Lambda_c^+\to \Sigma^+ \phi$
    are predicted to exhibit a large up-down asymmetry but a relatively small longitudinal asymmetry, suggesting significant contributions from the $A_2$ and $B_2$ amplitudes.
    
    \item In contrast, modes such as 
    $\Lambda_c^+\to \Sigma^0 \rho^+, \Lambda_c^+\to\Sigma^+\rho^0$
     are predicted to have a small $\alpha$ but large $P_L$. This implies that $\mathrm{Re}[(S+D)^* P_1]$ and $\mathrm{Re}(S^*P_2)$ are comparable but opposite in signs. From Eq. \eqref{eq:pw}, one can infer that $\mathrm{Re}(S^*P_2) \propto -A_1 B_1$ and $\mathrm{Re}[(S+D)^* P_1] \propto A_1 B_1$ if $A_2$ and $B_2$ are negligible. Therefore, decays with this type of decay asymmetry will have a small tensor coupling.
     
    \item In addition, sizable decay asymmetries, $\alpha$ and $P_L$, are found in certain channels, such as $\Lambda_c^+ \to p\,\overline{K}^{*0}$ and
     $\Lambda_c^+\to \Lambda\rho^+$.
      This pattern can be interpreted as a result of a suppressed interference term $\mathrm{Re}(S^* P_2)$, which in turn suggests the smallness of either the $A_1$ or $B_1$ form factor.
\end{enumerate}

It is interesting to note from Table \ref{tab:numCF} that
\begin{equation}
\mathcal{B}(\Lambda_c^+ \to \Sigma^0 \rho^+) \approx 
\mathcal{B}(\Lambda_c^+ \to \Sigma^+ \rho^0) = (4.6 \pm 1.5) \times 10^{-3},
\end{equation}
which  preserves isospin symmetry and satisfies the relation observed in Eq. \eqref{eq:sym}.
Additionally, we see that
\begin{equation}
\mathcal{B} (\Lambda_c^+ \to \Xi^0 K^{*+}) = (2.8 \pm 0.9) \times 10^{-3}, \quad
\mathcal{B} (\Xi_c^0 \to \Sigma^+ K^{*-}) = (5.1\pm 1.3) \times 10^{-3}.
\end{equation}
At first sight, it appears that the U-spin symmetry presented in Eq. \eqref{eq:sym} is broken. However, 
the kinetic factors involved in Eq. (\ref{eq:width}) lead to a rate of $\Xi_c^0 \to \Sigma^+ K^{*-}$ larger than that of $\Lambda_c^+ \to \Xi^0 K^{*+}$ by roughly a factor of 2, even though they are the same at the amplitude level.

\begin{table}[htbp]
\centering
\caption{Predictions of observables in singly Cabibbo-suppressed decays. }
\label{tab:numSCS}
\setlength{\tabcolsep}{8pt}
\begin{tabular}{l rrrr}
\toprule
{\rm Mode}&{\rm $10^{3}\mathcal{B}$}~~~~ &{$\alpha$}~~~~&	{$P_L$}~~~~	&{$\alpha_V$}~~~~\\
\midrule

$	\Lambda_c^+	\to	\Lambda	K^{*+}	$	&	$1.29\pm0.44$	&	$0.39 ^{+0.41}_{-0.33}$	&	$-0.45 ^{+0.28}_{-0.37}$	&	$-0.90\pm1.38$	\\
$	\Lambda_c^+	\to	\Sigma^0	K^{*+}	$	&	$0.23\pm0.11$	&	$-0.41 ^{+0.45}_{-0.36}$	&	$0.35\pm0.25$	&	$-0.49\pm1.83$	\\
$	\Lambda_c^+	\to	\Sigma^+	K^{*0}	$	&	$0.47\pm0.23$	&	$-0.41 ^{+0.45}_{-0.36}$	&	$0.35\pm0.25$	&	$-0.49\pm1.84$	\\
$	\Lambda_c^+	\to	p	\rho^0  	$	&	$0.93\pm0.17$	&	$-0.37 ^{+0.18}_{-0.52}$	&	$-0.41\pm0.55$	&	$10.80 ^{+23.22}_{-6.53}$	\\
$	\Lambda_c^+	\to	p	\phi  	$	&	$0.95\pm0.13$	&	$0.13 ^{+0.29}_{-0.56}$	&	$-0.70\pm0.46$	&	$1.51\pm9.58$	\\
$	\Lambda_c^+	\to	p	\omega	$	&	$0.90\pm0.10$	&	$0.84 ^{+0.08}_{-0.56}$	&	$0.91\pm0.42$	&	$6.51 ^{+16.32}_{-5.36}$	\\
$	\Lambda_c^+	\to	n	\rho^+	$	&	$5.04\pm3.19$	&	$0.47 ^{+0.24}_{-0.56}$	&	$-0.32 ^{+0.06}_{-0.40}$	&	$0.12\pm3.37$	\\
$	\Xi_c^0	\to	\Lambda	\rho^0  	$	&	$0.21\pm0.11$	&	$0.09 ^{+0.45}_{-0.52}$	&	$-0.03\pm0.36$	&	$-0.85\pm1.09$	\\
$	\Xi_c^0	\to	\Lambda	\phi  	$	&	$0.63\pm0.08$	&	$-0.30 ^{+0.41}_{-0.27}$	&	$-0.32\pm0.57$	&	$5.16\pm26.09$	\\
$	\Xi_c^0	\to	\Lambda	\omega	$	&	$1.24\pm0.24$	&	$0.80 ^{+0.15}_{-0.25}$	&	$0.65\pm0.59$	&	$7.47\pm25.07$	\\
$	\Xi_c^0	\to	\Sigma^0	\rho^0	$	&	$0.48\pm0.07$	&	$-0.86 ^{+0.29}_{-0.10}$	&	$-0.73\pm0.25$	&	$16.15 ^{+43.07}_{-10.88}$	\\
$	\Xi_c^0	\to	\Sigma^0	\phi   	$	&	$0.30\pm0.07$	&	$0.29 ^{+0.18}_{-0.63}$	&	$0.43\pm0.49$	&	$6.81 ^{+19.57}_{-5.26}$	\\
$	\Xi_c^0	\to	\Sigma^0	\omega	$	&	$0.67\pm0.24$	&	$-0.05 ^{+0.36}_{-0.42}$	&	$-0.62\pm0.54$	&	$3.58\pm18.96$	\\
$	\Xi_c^0	\to	\Sigma^+	\rho_-	$	&	$0.36\pm0.10$	&	$-0.24 ^{+0.41}_{-0.40}$	&	$-0.48\pm0.32$	&	$6.09\pm24.64$	\\
$	\Xi_c^0	\to	\Sigma^-	\rho^+	$	&	$2.66\pm1.01$	&	$0.87 ^{+0.06}_{-0.59}$	&	$-0.93\pm0.95$	&	$-0.94\pm1.86$	\\
$	\Xi_c^0	\to	\Xi^0	K^{*0}	$	&	$0.32\pm0.11$	&	$0.04 ^{+0.43}_{-0.84}$	&	$0.89\pm0.24$	&	$1.38\pm10.12$	\\
$	\Xi_c^0	\to	\Xi^-	K^{*+}	$	&	$2.06\pm0.43$	&	$0.89 ^{+0.06}_{-0.48}$	&	$-0.93\pm0.97$	&	$-0.96\pm1.34$	\\
$	\Xi_c^0	\to	p	K^{*-}	$	&	$0.34\pm0.09$	&	$-0.23 ^{+0.41}_{-0.40}$	&	$-0.48\pm0.32$	&	$5.94\pm24.10$	\\
$	\Xi_c^0	\to	n	\overline{K}^{*0} 	$	&	$0.47\pm0.17$	&	$0.18 ^{+0.38}_{-0.94}$	&	$0.88\pm0.25$	&	$2.22\pm13.71$	\\
$	\Xi_c^+	\to	\Lambda	\rho^+	$	&	$1.85\pm2.28$	&	$-0.72 ^{+0.88}_{-0.08}$	&	$0.38 ^{+0.43}_{-0.58}$	&	$-0.49\pm7.53$	\\
$	\Xi_c^+	\to	\Sigma^0	\rho^+	$	&	$3.12\pm1.50$	&	$0.78 ^{+0.14}_{-0.45}$	&	$-0.74 ^{+0.34}_{-0.19}$	&	$-0.89\pm1.32$	\\
$	\Xi_c^+	\to	\Sigma^+	\rho^0	$	&	$0.57\pm0.32$	&	$0.19 ^{+0.32}_{-0.74}$	&	$-0.78\pm0.42$	&	$0.85\pm6.51$	\\
$	\Xi_c^+	\to	\Sigma^+	\phi	$	&	$4.42\pm3.58$	&	$0.28 ^{+0.23}_{-0.60}$	&	$0.33\pm0.90$	&	$19.57 ^{+35.67}_{-11.75}$	\\
$	\Xi_c^+	\to	\Sigma^+	\omega	$	&	$4.03\pm1.45$	&	$-0.05 ^{+0.36}_{-0.42}$	&	$-0.62\pm0.54$	&	$3.58\pm18.96$	\\
$	\Xi_c^+	\to	\Xi^0	K^{*+}	$	&	$9.54\pm4.96$	&	$0.49 ^{+0.22}_{-0.51}$	&	$-0.35 ^{+0.07}_{-0.36}$	&	$-0.08\pm2.76$	\\
$	\Xi_c^+	\to	p	\overline{K}^{*0} 	$	&	$1.66\pm0.75$	&	$-0.39 ^{+0.48}_{-0.36}$	&	$0.30\pm0.12$	&	$-0.26\pm2.68$	\\

\hline
\hline
\end{tabular}
\end{table}

\begin{table}[t]
\centering
\caption{Predictions of observables in doubly Cabibbo-suppressed decays. }
\label{tab:numDCS}
\setlength{\tabcolsep}{12pt}
\begin{tabular}{l rrrr}
\toprule
{\rm Mode}&{\rm $10^{4}\mathcal{B}$}~~~~ &{$\alpha$}~~~~&	{$P_L$}~~~~	&{$\alpha_V$}~~~~\\
\midrule

$	\Lambda_c^+	\to	p	K^{*0}	$	&	$0.42\pm0.12$	&	$0.03 ^{+0.28}_{-0.78}$	&	$-0.40\pm0.48$	&	$0.75\pm5.02$	\\
$	\Lambda_c^+	\to	n	K^{*+}	$	&	$1.38\pm0.49$	&	$0.36 ^{+0.36}_{-0.33}$	&	$-0.31 ^{+0.10}_{-0.29}$	&	$-0.95\pm1.98$	\\
$	\Xi_c^0	\to	\Lambda	K^{*0}	$	&	$0.23\pm0.05$	&	$-0.08 ^{+0.24}_{-0.76}$	&	$0.12\pm0.56$	&	$7.63\pm33.34$	\\
$	\Xi_c^0	\to	\Sigma^0	K^{*0}	$	&	$0.15\pm0.01$	&	$-0.87 ^{+0.23}_{-0.08}$	&	$-0.84\pm0.22$	&	$13.04 ^{+26.66}_{-8.69}$	\\
$	\Xi_c^0	\to	\Sigma^-	K^{*+}	$	&	$1.24\pm0.31$	&	$0.88 ^{+0.06}_{-0.52}$	&	$-0.93\pm0.97$	&	$-0.95\pm1.51$	\\
$	\Xi_c^0	\to	p	\rho^-	$	&	$0.23\pm0.08$	&	$-0.27 ^{+0.40}_{-0.39}$	&	$-0.48\pm0.36$	&	$7.60\pm29.87$	\\
$	\Xi_c^0	\to	n	\rho^0	$	&	$0.12\pm0.04$	&	$-0.27 ^{+0.40}_{-0.39}$	&	$-0.48\pm0.36$	&	$7.59\pm29.85$	\\
$	\Xi_c^0	\to	n	\phi	$	&	$0.36\pm0.07$	&	$0.33 ^{+0.39}_{-0.55}$	&	$0.82\pm0.29$	&	$3.80\pm18.48$	\\
$	\Xi_c^0	\to	n	\omega	$	&	$0.38\pm0.24$	&	$0.79 ^{+0.11}_{-0.56}$	&	$0.08\pm0.98$	&	$2.61\pm11.45$	\\
$	\Xi_c^+	\to	\Lambda	K^{*+}	$	&	$21.16\pm10.39$	&	$0.30 ^{+0.24}_{-0.48}$	&	$0.01 ^{+0.06}_{-0.40}$	&	$-0.24\pm4.03$	\\
$	\Xi_c^+	\to	\Sigma^0	K^{*+}	$	&	$1.86\pm0.47$	&	$0.88 ^{+0.06}_{-0.52}$	&	$-0.93\pm0.97$	&	$-0.95\pm1.51$	\\
$	\Xi_c^+	\to	\Sigma^+	K^{*0}	$	&	$0.93\pm0.05$	&	$-0.87 ^{+0.23}_{-0.08}$	&	$-0.84\pm0.22$	&	$12.75 ^{+26.66}_{-8.69}$	\\
$	\Xi_c^+	\to	p	\rho^0	$	&	$0.35\pm0.12$	&	$-0.27 ^{+0.40}_{-0.39}$	&	$-0.48\pm0.36$	&	$7.58\pm29.80$	\\
$	\Xi_c^+	\to	p	\phi	$	&	$1.08\pm0.22$	&	$0.33 ^{+0.39}_{-0.55}$	&	$0.82\pm0.29$	&	$3.79\pm18.45$	\\
$	\Xi_c^+	\to	p	\omega	$	&	$1.15\pm0.70$	&	$0.79 ^{+0.11}_{-0.56}$	&	$0.08\pm0.98$	&	$2.61\pm11.43$	\\
$	\Xi_c^+	\to	n	\rho^+	$	&	$0.70\pm0.25$	&	$-0.27 ^{+0.40}_{-0.39}$	&	$-0.48\pm0.36$	&	$7.57\pm29.78$	\\

\hline
\end{tabular}
\end{table}

We also compare our predictions with experimental measurements and summarize the results in Table \ref{tab:comp1}. Among the 24 measured observables, there are 19 branching fractions and 5 up-down asymmetries. Our predictions for the up-down asymmetries are in excellent agreement with all the measured values. As for the branching fractions, all agree with experimental measurements, except for three modes: $\Lambda_c^+ \to \Sigma^+ K^{*0}$, $\Lambda_c^+ \to \Sigma^0 K^{*+}$, and $\Xi_c^+ \to p \overline{K}^{*0}$. These three modes are governed by the topological diagram $\tilde{C}'=C'-2E_{1S}$ with the  amplitude relations shown in Eq. \eqref{eq:sym}.

The measurements of $\Lambda_c^+ \to \Sigma^+ K^{*0}$ and $\Xi_c^+ \to p \overline{K}^{*0}$ were carried out by FOCUS nearly two decades ago, yielding $\mathcal{B}(\Lambda_c^+ \to \Sigma^+ K^{*0}) = (3.5 \pm 1.0) \times 10^{-3}$~\cite{FOCUS:2002rvb} and $\mathcal{B}(\Xi_c^+ \to p \overline{K}^{*0}) = (3.3 \pm 1.7) \times 10^{-3}$~\cite{FOCUS:2001ovr}. In contrast, $\Lambda_c^+ \to \Sigma^0 K^{*+}$ was recently measured by BESIII in 2026, yielding $\mathcal{B}(\Lambda_c^+ \to \Sigma^0 K^{*+}) = (1.23 \pm 0.57) \times 10^{-3}$~\cite{BESIII:2026vjj}. 
It appears the measured values obtained by BESIII and FOCUS satisfy the relation
presented in Eq. \eqref{eq:sym}.
Such an isospin relation is also respected in our theoretical predictions.
However, 
our prediction for $\Lambda_c^+ \to \Sigma^0 K^{*+}$ deviates from the BESIII measurement by approximately {\color{blue}{$4 \sigma$}}, while the predicted ${\cal B}(\Lambda_c^+ \to \Sigma^+ K^{*0})$ deviates from the FOCUS measurement in 2002 by 
more than
 $5 \sigma$.  
 On the other hand, the prediction for $\Xi_c^+ \to p \overline{K}^{*0}$ is consistent with experimental measured value.
Interestingly, a similar behavior for $\Lambda_c^+ \to \Sigma^+ K^{*0}$ and $\Lambda_c^+ \to \Sigma^0 K^{*+}$ is also observed by another theory group~\cite{Geng:2020zgr}, as shown in Table \ref{tab:comp2}.
We expect that future experiments will help resolve the discrepancies between theory and experiment for these three modes.

\begin{table}[htbp]
\centering
\caption{The branching fractions and decay asymmetries of the $\mathcal{B}_c \to \mathcal{B} V$ decays,
 where the order of magnitude of the branching fractions in the first and second rows is  $10^{-2}$ and $10^{-3}$, respectively. 
 }
\label{tab:comp1}
\setlength{\tabcolsep}{14pt}
\begin{tabular}{l l r r r  }
\hline
{\rm Mode} & $\mathcal{B}_{\rm theo}$ & $\mathcal{B}_{\rm exp}$~~~& $P_{L,{\rm{theo}}}$ & $P_{L,{\rm{exp}}}$\\

\hline

$\Lambda^+_c \to p \overline{K}^{*0}$ & $	1.40	\pm	0.07	$ & $1.40\pm0.07$& $-0.84 \pm 0.21$ &$-0.835\pm0.215$ \\
$\Lambda^+_c \to \Lambda \rho^+$ & $	4.10	\pm	0.50	$ & $4.10\pm0.50$& $-0.76\pm0.07$ & $-0.76\pm0.07$\\
$\Lambda^+_c \to \Sigma^+\rho^0$ & $	0.46	\pm	0.15	$ & $<1.70$ & &\\
$\Lambda^+_c \to \Sigma^+\omega$ & $	1.67	\pm	0.17	$ & $1.72\pm0.20$ & & \\
$\Xi^+_c \to \Sigma^+ \overline{K}^{*0}$ & $	3.58  \pm	1.01	$ & $2.30\pm1.10$& & \\
\hline				
$\Lambda^+_c \to p\phi$ & $	0.95	\pm	0.13	$ & $1.05\pm0.14$& & \\
$\Lambda^+_c \to \Sigma^+\phi$ & $	4.09	\pm	0.49	$ & $4.00\pm0.50$ & &  \\
$\Lambda^+_c \to \Sigma^+ K^{*0}$ & $	0.47	\pm	0.23	$ & $3.50\pm1.00$ & &\\
$\Lambda^+_c \to \Sigma^0 K^{*+}$ & $	0.23	\pm	0.11	$ & $1.23\pm0.57$& & \\
				
$\Lambda^+_c \to p\rho^0$ & $	0.93	\pm	0.17	$ & $1.50\pm0.40$ & &  \\
$\Lambda^+_c \to \Lambda K^{*+}$ & $	1.29	\pm	0.44	$ & $1.29\pm0.44$& &  \\
$\Lambda^+_c \to p\omega$ & $	0.90	\pm	0.10	$ & $0.90\pm0.10$& & \\
$\Xi^0_c \to \Lambda\overline{K}^{*0}$ & $	2.90	\pm	0.57	$ & $2.60\pm0.60$& $0.16 \pm 0.22$ & $0.15\pm0.22$ \\
$\Xi^0_c \to \Sigma^0 \overline{K}^{*0}$ & $	10.15	\pm	1.66	$ & $9.90\pm1.90$& & \\
$\Xi^0_c \to \Sigma^+K^{*-}$ & $	5.15	\pm	1.26	$ & $4.90\pm1.30$& $-0.48 \pm 0.30$ & $-0.50\pm0.30$\\
$\Xi^0_c \to \Lambda\phi$ & $	0.63 	\pm	0.08	$ & $0.49\pm0.13$& & \\
$\Xi^+_c \to p\phi$ & $	0.11	\pm	0.02	$ & $0.12\pm0.06$& & \\
$\Xi^+_c \to p\overline{K}^{*0}$ & $	1.66	\pm	0.75	$ & $3.30\pm1.70$& $ 0.30 \pm 0.12$ &$0.303 \pm 0.121$ \\
$\Xi^+_c \to \Sigma^+\phi$ & $	1.79	\pm	0.00	$ & $<3.20$ & &\\

\hline
\end{tabular}
\end{table}

\begin{table}[htbp]
    \centering
    \caption{
    Branching fractions (first row in units of 
$10^{-2}$, second row in units of $10^{-3}$) and the decay asymmetries $\alpha$ for the $\mathcal{B}_c \to \mathcal{B} V$
decays, where the upper entries correspond to the branching fractions and the lower entries to the decay asymmetries.
    }
    \label{tab:comp2}
    \begin{tabular}{ l r r r r r} \hline
    Mode & This work & Geng \cite{Geng:2020zgr} & ~~Jia \cite{Jia:2024pyb}~~ & Hsiao \cite{Hsiao:2019yur} & Exp \\ \hline
    
  \multirow{2}{*}{$\Lambda_c^+ \rightarrow \Lambda\rho^{+}$}
    & $4.10 \pm 0.50$ & $4.81 \pm 0.58$ & $6.26^{+2.44}_{-1.39}$ & $0.74 \pm 0.34$ & $4.1 \pm 0.5$ \\
    & $-0.76\pm0.07$& $0.93 \pm 0.05$ & & & $-0.76 \pm 0.07$ \\
    \multirow{2}{*}{$\Lambda_c^+ \rightarrow p \overline{K}^{*0}$}
    & $1.40 \pm 0.07$ & $2.03 \pm 0.25$ & $3.70^{+1.29}_{-3.39}$ & $1.90 \pm 0.30$ & $1.4 \pm 0.1$ \\
    & $-0.84 \pm 0.22$ & $-0.18 \pm 0.05$ &  &  & $-0.84 \pm 0.22$ \\
    \multirow{1}{*}{$\Lambda_c^+ \rightarrow \Sigma^{+}\rho^{0}$}
    & $0.46 \pm 0.15$ & $1.43 \pm 0.42$ & $0.77^{+1.38}_{-0.53}$ & $0.61 \pm 0.46$ & $< 1.7$ \\
    \multirow{1}{*}{$\Lambda_c^+ \rightarrow \Sigma^{+}\omega$}
    & $1.67 \pm 0.17$ & $1.81 \pm 0.19$ & $2.06^{+0.40}_{-1.78}$ & $1.60 \pm 0.70$ & $1.7\pm 0.2$ \\
    \multirow{1}{*}{$\Xi_c^+ \rightarrow \Sigma^{+}\overline{K}^{*0}$}
    & $3.58 \pm 1.01$ & $1.40 \pm 0.69$ & & $10.10 \pm 2.90$ & $2.3 \pm 1.1$ \\
    \hline
    \multirow{1}{*}{$\Lambda_c^+ \rightarrow \Sigma^{+}\phi$}
    & $4.09 \pm 0.49$ & $3.90 \pm 0.60$ & $3.30^{+0.80}_{-2.90}$ & $3.90 \pm 0.60$ & $4.0 \pm 0.5$ \\
    \multirow{1}{*}{$\Lambda_c^+ \rightarrow p\phi$}
    & $0.95 \pm 0.13$ & $0.87 \pm 0.14$ & $1.37^{+1.13}_{-0.65}$ & $1.04 \pm 0.21$ & $1.1 \pm 0.1$ \\
    \multirow{1}{*}{$\Lambda_c^+ \rightarrow \Sigma^{+}K^{*0}$}
    & $0.47 \pm 0.23$ & $0.38 \pm 0.09$ & $2.10^{+1.37}_{-0.86}$ & $2.30 \pm 0.60$ & $3.5 \pm 1.0$ \\
     \multirow{1}{*}{$\Lambda_c^+ \rightarrow \Sigma^0 K^{*+}$}
    & $0.23 \pm 0.11$ & $0.18 \pm 0.04$ & $1.60^{+0.89}_{-0.62}$ & $1.2 \pm 0.3$ & $1.23 \pm 0.57$ \\
    \multirow{1}{*}{$\Lambda_c^+ \rightarrow p\rho^{0}$}
    & $0.93 \pm 0.17$ & $0.02^{+0.07}_{-0.02}$ & $2.72^{+1.27}_{-1.87}$ & $0.35 \pm 0.29$ & $1.5 \pm 0.4$ \\
    \multirow{1}{*}{$\Lambda_c^+ \rightarrow \Lambda K^{*+}$}
    & $1.29 \pm 0.44$ & $3.35 \pm 0.37$ & $4.71^{+0.48}_{-0.20}$ & $2.00 \pm 0.50$ & $2.4\pm 0.6$ \\
    \multirow{1}{*}{$\Lambda_c^+ \rightarrow p \omega$}
    & $0.90 \pm 0.10$ & $0.63 \pm 0.34$ & $1.26^{+0.45}_{-0.37}$ & $1.14 \pm 0.54$ & $0.9 \pm 0.1$ \\
    \multirow{2}{*}{$\Xi_c^+ \rightarrow p \overline{K}^{*0}$}
    & $1.66 \pm 0.75$ & $4.71 \pm 1.22$ & & $7.80 \pm 2.20$ & $3.3 \pm 1.7$ \\
    & $0.30 \pm 0.12$ & $-0.12 \pm 0.15$ & &  & $0.303 \pm 0.121$ \\
    \multirow{1}{*}{$\Xi_c^+ \rightarrow \Sigma^{+}\phi$}
    & $1.79 \pm 0.00$ & $1.82 \pm 0.40$ & & $1.90 \pm 0.90$ & $< 3.2$ \\
    \multirow{1}{*}{$\Xi_c^+ \rightarrow p \phi$}
    & $0.11 \pm 0.02$ & $0.23 \pm 0.04$ &  & $0.15 \pm 0.07$ & $0.12 \pm 0.06$ \\
    \multirow{2}{*}{$\Xi_c^0 \rightarrow \Lambda \overline{K}^{*0}$}
    & $2.90 \pm 0.57$ & $13.70 \pm 2.60$ & & $4.60 \pm 2.10$ & $2.6 \pm 0.6$ \\
    & $0.16 \pm 0.22$ & $-0.28\pm 0.10$ & & & $0.15 \pm 0.22$ \\
    \multirow{1}{*}{$\Xi_c^0 \rightarrow \Sigma^{0}\overline{K}^{*0}$}
    & $10.15 \pm 1.66$ & $4.20 \pm 2.30$ & & $2.70 \pm 2.20$ & $9.9 \pm 1.9$ \\ 
    \multirow{2}{*}{$\Xi_c^0 \rightarrow \Sigma^{+}K^{*-}$}
    & $5.15 \pm 1.26$ & $2.40 \pm 1.70$ & & $9.30 \pm 2.90$ & $4.9 \pm 1.3$ \\
    & $-0.48 \pm 0.30$ & $-0.37\pm 0.31$ & & & $-0.50\pm 0.30$ \\
    \multirow{1}{*}{$\Xi^0_c \rightarrow \Lambda\phi$}
    & $0.63 \pm 0.08$ & $0.44 \pm 0.08$ &  & $0.84 \pm 0.39$ & $0.49 \pm 0.13$ \\
    
    \hline
    \end{tabular}
    \end{table}


\section{Conclusions}
\label{sec:con}

In this work, we have further advanced the application of the topological diagram approach (TDA) to weak decays of charmed baryons. By incorporating the KPW theorem, we find that only five independent sets of topological amplitudes are required. This result is consistent with the case of $\mathcal{B}_c \to \mathcal{B} P$ decays. Compared with $\mathcal{B}_c \to \mathcal{B} P$, however, the processes $\mathcal{B}_c \to \mathcal{B} V$ involve much richer polarization information, which leads to additional complexity in the analysis.
It is well known that there are vector- and tensor-type couplings for the $\B\B V$ interaction. In this work we incorporate both contributions in terms of four independent form factors. Relations among different decay channels at the amplitude level, arising from isospin, U-spin, and V-spin symmetries, are explicitly derived. We perform global fits in terms of both partial waves and form factors. Physical observables, including branching fractions, up-down decay asymmetries, longitudinal polarizations, and polarization observables in subsequent decays, are systematically calculated.
Our numerical results can be summarized as follows:
\begin{itemize}

\item The form factors $A_2$ and $B_2$ are found to be comparable in magnitude to $A_1$ and $B_1$, as shown in Table~\ref{tab:fitpara_2}. This indicates that their contributions cannot be neglected and further implies that the tensor coupling in the $\B\B V$ interaction plays an important role in $\mathcal{B}_c \to \mathcal{B} V$ decays.

\item The partial-wave contributions, including the parity-violating $S$- and $D$-waves as well as the spin-flipped and spin-nonflipped $P$-waves, associated with each topological diagram, have been explicitly extracted from the experimental data.

\item Branching fractions as well as other observables for all CF, SCS, and DCS two-body weak decays of charmed baryons into an octet baryon and a vector meson have been calculated. Except for the 
two modes $\Lambda_c^+\to\Sigma^+ K^{*0}$ and $\Lambda_c^+\to\Sigma^0 K^{*+}$, 
our predictions for all measured channels are in good agreement with the available experimental data.

\item Comparisons with predictions from other theoretical approaches and groups have also been presented.

\end{itemize}

With the continuous accumulation of data from flavor experiments, it is highly promising that more measurements will become available in the near future. These data will provide crucial tests of the predictions of the TDA framework and help clarify those channels in which current theoretical predictions deviate from existing experimental measurements.

\acknowledgments
We are grateful to Mingyue Jia, Qiaoyi Wen,  
{Cai-Ping Jia and Peirong Li
for valuable discussions.
This research was supported in part by the
 National Natural
Science Foundation of China under Grant No. 12475095
and National Science and Technology Council of R.O.C. under
Grant No. 114-2112-M-001-039.



\normalem
\clearpage
\bibliographystyle{modified-apsrev4-2}
\bibliography{reference}

\begin{thebibliography}{30}%
\makeatletter
\providecommand \@ifxundefined [1]{%
 \@ifx{#1\undefined}
}%
\providecommand \@ifnum [1]{%
 \ifnum #1\expandafter \@firstoftwo
 \else \expandafter \@secondoftwo
 \fi
}%
\providecommand \@ifx [1]{%
 \ifx #1\expandafter \@firstoftwo
 \else \expandafter \@secondoftwo
 \fi
}%
\providecommand \natexlab [1]{#1}%
\providecommand \enquote  [1]{``#1''}%
\providecommand \bibnamefont  [1]{#1}%
\providecommand \bibfnamefont [1]{#1}%
\providecommand \citenamefont [1]{#1}%
\providecommand \href@noop [0]{\@secondoftwo}%
\providecommand \href [0]{\begingroup \@sanitize@url \@href}%
\providecommand \@href[1]{\@@startlink{#1}\@@href}%
\providecommand \@@href[1]{\endgroup#1\@@endlink}%
\providecommand \@sanitize@url [0]{\catcode `\\12\catcode `\$12\catcode
  `\&12\catcode `\#12\catcode `\^12\catcode `\_12\catcode `\%12\relax}%
\providecommand \@@startlink[1]{}%
\providecommand \@@endlink[0]{}%
\providecommand \url  [0]{\begingroup\@sanitize@url \@url }%
\providecommand \@url [1]{\endgroup\@href {#1}{\urlprefix }}%
\providecommand \urlprefix  [0]{URL }%
\providecommand \Eprint [0]{\href }%
\providecommand \doibase [0]{https://doi.org/}%
\providecommand \selectlanguage [0]{\@gobble}%
\providecommand \bibinfo  [0]{\@secondoftwo}%
\providecommand \bibfield  [0]{\@secondoftwo}%
\providecommand \translation [1]{[#1]}%
\providecommand \BibitemOpen [0]{}%
\providecommand \bibitemStop [0]{}%
\providecommand \bibitemNoStop [0]{.\EOS\space}%
\providecommand \EOS [0]{\spacefactor3000\relax}%
\providecommand \BibitemShut  [1]{\csname bibitem#1\endcsname}%
\let\auto@bib@innerbib\@empty
\bibitem [{\citenamefont {Li}\ \emph {et~al.}(2026)\citenamefont {Li},
  \citenamefont {Lyu},\ and\ \citenamefont {Zheng}}]{Li:2025nzx}%
  \BibitemOpen
  \bibfield  {author} {\bibinfo {author} {\bibfnamefont {P.-R.}\ \bibnamefont
  {Li}}, \bibinfo {author} {\bibfnamefont {X.-R.}\ \bibnamefont {Lyu}},\ and\
  \bibinfo {author} {\bibfnamefont {Y.}~\bibnamefont {Zheng}},\ }\bibinfo
  {title} {{Experimental overview on the charmed baryon decays}},\ \href
  {https://doi.org/10.1088/1674-1137/ae1187} {\bibfield  {journal} {\bibinfo
  {journal} {Chin. Phys.}\ }\textbf {\bibinfo {volume} {50}},\ \bibinfo {pages}
  {022002} (\bibinfo {year} {2026})},\ \Eprint
  {https://arxiv.org/abs/2509.19141} {arXiv:2509.19141 [hep-ex]} \BibitemShut
  {NoStop}%
\bibitem [{\citenamefont {Cheng}(2022)}]{Cheng:2021qpd}%
  \BibitemOpen
  \bibfield  {author} {\bibinfo {author} {\bibfnamefont {H.-Y.}\ \bibnamefont
  {Cheng}},\ }\bibinfo {title} {{Charmed baryon physics circa 2021}},\ \href
  {https://doi.org/10.1016/j.cjph.2022.06.021} {\bibfield  {journal} {\bibinfo
  {journal} {Chin. J. Phys.}\ }\textbf {\bibinfo {volume} {78}},\ \bibinfo
  {pages} {324} (\bibinfo {year} {2022})},\ \Eprint
  {https://arxiv.org/abs/2109.01216} {arXiv:2109.01216 [hep-ph]} \BibitemShut
  {NoStop}%
\bibitem [{\citenamefont {Zhong}\ \emph
  {et~al.}(2024{\natexlab{a}})\citenamefont {Zhong}, \citenamefont {Xu},\ and\
  \citenamefont {Cheng}}]{Zhong:2024zme}%
  \BibitemOpen
  \bibfield  {author} {\bibinfo {author} {\bibfnamefont {H.}~\bibnamefont
  {Zhong}}, \bibinfo {author} {\bibfnamefont {F.}~\bibnamefont {Xu}},\ and\
  \bibinfo {author} {\bibfnamefont {H.-Y.}\ \bibnamefont {Cheng}},\ }\bibinfo
  {title} {{Topological Diagrams and Hadronic Weak Decays of Charmed
  Baryons}},\ \href@noop {} {\  (\bibinfo {year} {2024}{\natexlab{a}})},\
  \Eprint {https://arxiv.org/abs/2401.15926} {arXiv:2401.15926 [hep-ph]}
  \BibitemShut {NoStop}%
\bibitem [{\citenamefont {Zhong}\ \emph
  {et~al.}(2024{\natexlab{b}})\citenamefont {Zhong}, \citenamefont {Xu},\ and\
  \citenamefont {Cheng}}]{Zhong:2024qqs}%
  \BibitemOpen
  \bibfield  {author} {\bibinfo {author} {\bibfnamefont {H.}~\bibnamefont
  {Zhong}}, \bibinfo {author} {\bibfnamefont {F.}~\bibnamefont {Xu}},\ and\
  \bibinfo {author} {\bibfnamefont {H.-Y.}\ \bibnamefont {Cheng}},\ }\bibinfo
  {title} {{Analysis of hadronic weak decays of charmed baryons in the
  topological diagrammatic approach}},\ \href
  {https://doi.org/10.1103/PhysRevD.109.114027} {\bibfield  {journal} {\bibinfo
   {journal} {Phys. Rev. D}\ }\textbf {\bibinfo {volume} {109}},\ \bibinfo
  {pages} {114027} (\bibinfo {year} {2024}{\natexlab{b}})},\ \Eprint
  {https://arxiv.org/abs/2404.01350} {arXiv:2404.01350 [hep-ph]} \BibitemShut
  {NoStop}%
\bibitem [{\citenamefont {Cheng}\ \emph
  {et~al.}(2025{\natexlab{a}})\citenamefont {Cheng}, \citenamefont {Xu},\ and\
  \citenamefont {Zhong}}]{Cheng:2024lsn}%
  \BibitemOpen
  \bibfield  {author} {\bibinfo {author} {\bibfnamefont {H.-Y.}\ \bibnamefont
  {Cheng}}, \bibinfo {author} {\bibfnamefont {F.}~\bibnamefont {Xu}},\ and\
  \bibinfo {author} {\bibfnamefont {H.}~\bibnamefont {Zhong}},\ }\bibinfo
  {title} {{Hadronic weak decays of charmed baryons in the topological
  diagrammatic approach: An update}},\ \href
  {https://doi.org/10.1103/PhysRevD.111.034011} {\bibfield  {journal} {\bibinfo
   {journal} {Phys. Rev. D}\ }\textbf {\bibinfo {volume} {111}},\ \bibinfo
  {pages} {034011} (\bibinfo {year} {2025}{\natexlab{a}})},\ \Eprint
  {https://arxiv.org/abs/2410.04675} {arXiv:2410.04675 [hep-ph]} \BibitemShut
  {NoStop}%
\bibitem [{\citenamefont {Wang}(2026)}]{YHW:Beauty2026}%
  \BibitemOpen
  \bibfield  {author} {\bibinfo {author} {\bibfnamefont {Y.}~\bibnamefont
  {Wang}},\ }\href
  {https://indico.cern.ch/event/1595459/contributions/7058352/attachments/3286972/5876405/DHH_static.pdf}
  {\bibinfo {title} {Observation of the doubly charmed baryon
  {$\Omega_{cc}^{+}$}}},\ \bibinfo {howpublished} {Talk presented at the 21st
  International Conference on B-Physics at Frontier Machines (BEAUTY 2026),
  Maastricht, The Netherlands} (\bibinfo {year} {2026}),\ \bibinfo {note}
  {presentation slides}\BibitemShut {NoStop}%
\bibitem [{\citenamefont {Cheng}\ \emph {et~al.}(2020)\citenamefont {Cheng},
  \citenamefont {Meng}, \citenamefont {Xu},\ and\ \citenamefont
  {Zou}}]{Cheng:2020wmk}%
  \BibitemOpen
  \bibfield  {author} {\bibinfo {author} {\bibfnamefont {H.-Y.}\ \bibnamefont
  {Cheng}}, \bibinfo {author} {\bibfnamefont {G.}~\bibnamefont {Meng}},
  \bibinfo {author} {\bibfnamefont {F.}~\bibnamefont {Xu}},\ and\ \bibinfo
  {author} {\bibfnamefont {J.}~\bibnamefont {Zou}},\ }\bibinfo {title}
  {{Two-body weak decays of doubly charmed baryons}},\ \href
  {https://doi.org/10.1103/PhysRevD.101.034034} {\bibfield  {journal} {\bibinfo
   {journal} {Phys. Rev. D}\ }\textbf {\bibinfo {volume} {101}},\ \bibinfo
  {pages} {034034} (\bibinfo {year} {2020})},\ \Eprint
  {https://arxiv.org/abs/2001.04553} {arXiv:2001.04553 [hep-ph]} \BibitemShut
  {NoStop}%
\bibitem [{\citenamefont {Pakvasa}\ \emph {et~al.}(1990)\citenamefont
  {Pakvasa}, \citenamefont {Rosen},\ and\ \citenamefont
  {Tuan}}]{Pakvasa:1990if}%
  \BibitemOpen
  \bibfield  {author} {\bibinfo {author} {\bibfnamefont {S.}~\bibnamefont
  {Pakvasa}}, \bibinfo {author} {\bibfnamefont {S.~P.}\ \bibnamefont {Rosen}},\
  and\ \bibinfo {author} {\bibfnamefont {S.~F.}\ \bibnamefont {Tuan}},\
  }\bibinfo {title} {{Parity Violation and Flavor Selection Rules in Charmed
  Baryon Decays}},\ \href {https://doi.org/10.1103/PhysRevD.42.3746} {\bibfield
   {journal} {\bibinfo  {journal} {Phys. Rev. D}\ }\textbf {\bibinfo {volume}
  {42}},\ \bibinfo {pages} {3746} (\bibinfo {year} {1990})}\BibitemShut
  {NoStop}%
\bibitem [{\citenamefont {Cheng}\ and\ \citenamefont
  {Tseng}(1992)}]{Cheng:1991sn}%
  \BibitemOpen
  \bibfield  {author} {\bibinfo {author} {\bibfnamefont {H.-Y.}\ \bibnamefont
  {Cheng}}\ and\ \bibinfo {author} {\bibfnamefont {B.}~\bibnamefont {Tseng}},\
  }\bibinfo {title} {{Nonleptonic weak decays of charmed baryons}},\ \href
  {https://doi.org/10.1103/PhysRevD.46.1042} {\bibfield  {journal} {\bibinfo
  {journal} {Phys. Rev. D}\ }\textbf {\bibinfo {volume} {46}},\ \bibinfo
  {pages} {1042} (\bibinfo {year} {1992})},\ \bibinfo {note} {[Erratum: Phys.
  Rev. D 55, 1697 (1997)]}\BibitemShut {NoStop}%
\bibitem [{\citenamefont {Hsiao}\ \emph {et~al.}(2019)\citenamefont {Hsiao},
  \citenamefont {Yao},\ and\ \citenamefont {Zhao}}]{Hsiao:2019yur}%
  \BibitemOpen
  \bibfield  {author} {\bibinfo {author} {\bibfnamefont {Y.~K.}\ \bibnamefont
  {Hsiao}}, \bibinfo {author} {\bibfnamefont {Y.}~\bibnamefont {Yao}},\ and\
  \bibinfo {author} {\bibfnamefont {H.~J.}\ \bibnamefont {Zhao}},\ }\bibinfo
  {title} {{Two-body charmed baryon decays involving vector meson with $SU(3)$
  flavor symmetry}},\ \href {https://doi.org/10.1016/j.physletb.2019.03.031}
  {\bibfield  {journal} {\bibinfo  {journal} {Phys. Lett. B}\ }\textbf
  {\bibinfo {volume} {792}},\ \bibinfo {pages} {35} (\bibinfo {year} {2019})},\
  \Eprint {https://arxiv.org/abs/1902.08783} {arXiv:1902.08783 [hep-ph]}
  \BibitemShut {NoStop}%
\bibitem [{\citenamefont {Geng}\ \emph {et~al.}(2020)\citenamefont {Geng},
  \citenamefont {Liu},\ and\ \citenamefont {Tsai}}]{Geng:2020zgr}%
  \BibitemOpen
  \bibfield  {author} {\bibinfo {author} {\bibfnamefont {C.~Q.}\ \bibnamefont
  {Geng}}, \bibinfo {author} {\bibfnamefont {C.-W.}\ \bibnamefont {Liu}},\ and\
  \bibinfo {author} {\bibfnamefont {T.-H.}\ \bibnamefont {Tsai}},\ }\bibinfo
  {title} {{Charmed Baryon Weak Decays with Vector Mesons}},\ \href
  {https://doi.org/10.1103/PhysRevD.101.053002} {\bibfield  {journal} {\bibinfo
   {journal} {Phys. Rev. D}\ }\textbf {\bibinfo {volume} {101}},\ \bibinfo
  {pages} {053002} (\bibinfo {year} {2020})},\ \Eprint
  {https://arxiv.org/abs/2001.05079} {arXiv:2001.05079 [hep-ph]} \BibitemShut
  {NoStop}%
\bibitem [{\citenamefont {Jia}\ \emph {et~al.}(2024)\citenamefont {Jia},
  \citenamefont {Jiang}, \citenamefont {Wang},\ and\ \citenamefont
  {Yu}}]{Jia:2024pyb}%
  \BibitemOpen
  \bibfield  {author} {\bibinfo {author} {\bibfnamefont {C.-P.}\ \bibnamefont
  {Jia}}, \bibinfo {author} {\bibfnamefont {H.-Y.}\ \bibnamefont {Jiang}},
  \bibinfo {author} {\bibfnamefont {J.-P.}\ \bibnamefont {Wang}},\ and\
  \bibinfo {author} {\bibfnamefont {F.-S.}\ \bibnamefont {Yu}},\ }\bibinfo
  {title} {{Final-state rescattering mechanism of charmed baryon decays}},\
  \href {https://doi.org/10.1007/JHEP11(2024)072} {\bibfield  {journal}
  {\bibinfo  {journal} {JHEP}\ }\textbf {\bibinfo {volume} {11}},\ \bibinfo
  {pages} {072}},\ \Eprint {https://arxiv.org/abs/2408.14959} {arXiv:2408.14959
  [hep-ph]} \BibitemShut {NoStop}%
\bibitem [{\citenamefont {Ablikim}\ \emph
  {et~al.}(2026{\natexlab{a}})\citenamefont {Ablikim} \emph
  {et~al.}}]{BESIII:2026vjj}%
  \BibitemOpen
  \bibfield  {author} {\bibinfo {author} {\bibfnamefont {M.}~\bibnamefont
  {Ablikim}} \emph {et~al.} (\bibinfo {collaboration} {BESIII}),\ }\bibinfo
  {title} {{Measurements of branching fractions of
  $\Lambda_{c}^{+}\to\Sigma^{0}K_{S}^{0} \pi^{+}$ and
  $\Lambda_{c}^{+}\to\Sigma^{0}K_{S}^{0}K^{+}$}},\ \href@noop {} {\  (\bibinfo
  {year} {2026}{\natexlab{a}})},\ \Eprint {https://arxiv.org/abs/2602.22754}
  {arXiv:2602.22754 [hep-ex]} \BibitemShut {NoStop}%
\bibitem [{\citenamefont {Wang}\ \emph {et~al.}(2017)\citenamefont {Wang},
  \citenamefont {Yu},\ and\ \citenamefont {Zhao}}]{Wang:2017mqp}%
  \BibitemOpen
  \bibfield  {author} {\bibinfo {author} {\bibfnamefont {W.}~\bibnamefont
  {Wang}}, \bibinfo {author} {\bibfnamefont {F.-S.}\ \bibnamefont {Yu}},\ and\
  \bibinfo {author} {\bibfnamefont {Z.-X.}\ \bibnamefont {Zhao}},\ }\bibinfo
  {title} {{Weak decays of doubly heavy baryons: the $1/2\rightarrow 1/2$
  case}},\ \href {https://doi.org/10.1140/epjc/s10052-017-5360-1} {\bibfield
  {journal} {\bibinfo  {journal} {Eur. Phys. J. C}\ }\textbf {\bibinfo {volume}
  {77}},\ \bibinfo {pages} {781} (\bibinfo {year} {2017})},\ \Eprint
  {https://arxiv.org/abs/1707.02834} {arXiv:1707.02834 [hep-ph]} \BibitemShut
  {NoStop}%
\bibitem [{\citenamefont {Cheng}\ \emph {et~al.}(2018)\citenamefont {Cheng},
  \citenamefont {Kang},\ and\ \citenamefont {Xu}}]{Cheng:2018hwl}%
  \BibitemOpen
  \bibfield  {author} {\bibinfo {author} {\bibfnamefont {H.-Y.}\ \bibnamefont
  {Cheng}}, \bibinfo {author} {\bibfnamefont {X.-W.}\ \bibnamefont {Kang}},\
  and\ \bibinfo {author} {\bibfnamefont {F.}~\bibnamefont {Xu}},\ }\bibinfo
  {title} {{Singly Cabibbo-suppressed hadronic decays of $\Lambda_c^+$}},\
  \href {https://doi.org/10.1103/PhysRevD.97.074028} {\bibfield  {journal}
  {\bibinfo  {journal} {Phys. Rev. D}\ }\textbf {\bibinfo {volume} {97}},\
  \bibinfo {pages} {074028} (\bibinfo {year} {2018})},\ \Eprint
  {https://arxiv.org/abs/1801.08625} {arXiv:1801.08625 [hep-ph]} \BibitemShut
  {NoStop}%
\bibitem [{\citenamefont {Navas}\ \emph {et~al.}(2024)\citenamefont {Navas}
  \emph {et~al.}}]{ParticleDataGroup:2024cfk}%
  \BibitemOpen
  \bibfield  {author} {\bibinfo {author} {\bibfnamefont {S.}~\bibnamefont
  {Navas}} \emph {et~al.} (\bibinfo {collaboration} {Particle Data Group}),\
  }\bibinfo {title} {{Review of particle physics}},\ \href
  {https://doi.org/10.1103/PhysRevD.110.030001} {\bibfield  {journal} {\bibinfo
   {journal} {Phys. Rev. D}\ }\textbf {\bibinfo {volume} {110}},\ \bibinfo
  {pages} {030001} (\bibinfo {year} {2024})}\BibitemShut {NoStop}%
\bibitem [{\citenamefont {Aaij}\ \emph
  {et~al.}(2025{\natexlab{a}})\citenamefont {Aaij} \emph
  {et~al.}}]{LHCb:2025oww}%
  \BibitemOpen
  \bibfield  {author} {\bibinfo {author} {\bibfnamefont {R.}~\bibnamefont
  {Aaij}} \emph {et~al.} (\bibinfo {collaboration} {LHCb}),\ }\bibinfo {title}
  {{Measurement of the $ {\varOmega}_c^0 $ and $ {\Xi}_c^0 $ baryon lifetimes
  using hadronic b-baryon decays}},\ \href
  {https://doi.org/10.1007/JHEP09(2025)157} {\bibfield  {journal} {\bibinfo
  {journal} {JHEP}\ }\textbf {\bibinfo {volume} {09}},\ \bibinfo {pages}
  {157}},\ \Eprint {https://arxiv.org/abs/2506.13334} {arXiv:2506.13334
  [hep-ex]} \BibitemShut {NoStop}%
\bibitem [{\citenamefont {Ablikim}\ \emph {et~al.}(2022)\citenamefont {Ablikim}
  \emph {et~al.}}]{BESIII:2022udq}%
  \BibitemOpen
  \bibfield  {author} {\bibinfo {author} {\bibfnamefont {M.}~\bibnamefont
  {Ablikim}} \emph {et~al.} (\bibinfo {collaboration} {BESIII}),\ }\bibinfo
  {title} {{Partial wave analysis of the charmed baryon hadronic decay $
  {\Lambda}_c^{+} ${\textrightarrow}
  {\ensuremath{\Lambda}}{\ensuremath{\pi}}$^{+}${\ensuremath{\pi}}$^{0}$}},\
  \href {https://doi.org/10.1007/JHEP12(2022)033} {\bibfield  {journal}
  {\bibinfo  {journal} {JHEP}\ }\textbf {\bibinfo {volume} {12}},\ \bibinfo
  {pages} {033}},\ \Eprint {https://arxiv.org/abs/2209.08464} {arXiv:2209.08464
  [hep-ex]} \BibitemShut {NoStop}%
\bibitem [{\citenamefont {Jia}\ \emph {et~al.}(2021)\citenamefont {Jia} \emph
  {et~al.}}]{Belle:2021zsy}%
  \BibitemOpen
  \bibfield  {author} {\bibinfo {author} {\bibfnamefont {S.}~\bibnamefont
  {Jia}} \emph {et~al.} (\bibinfo {collaboration} {Belle}),\ }\bibinfo {title}
  {{Measurements of branching fractions and asymmetry parameters of $\Xi^0_c\to
  \Lambda\bar K^{*0}$, $\Xi^0_c\to \Sigma^0\bar K^{*0}$, and $\Xi^0_c\to
  \Sigma^+K^{*-}$ decays at Belle}},\ \href
  {https://doi.org/10.1007/JHEP06(2021)160} {\bibfield  {journal} {\bibinfo
  {journal} {JHEP}\ }\textbf {\bibinfo {volume} {06}},\ \bibinfo {pages}
  {160}},\ \Eprint {https://arxiv.org/abs/2104.10361} {arXiv:2104.10361
  [hep-ex]} \BibitemShut {NoStop}%
\bibitem [{\citenamefont {Aliev}\ \emph {et~al.}(2009)\citenamefont {Aliev},
  \citenamefont {Ozpineci}, \citenamefont {Savci},\ and\ \citenamefont
  {Zamiralov}}]{Aliev:2009ei}%
  \BibitemOpen
  \bibfield  {author} {\bibinfo {author} {\bibfnamefont {T.~M.}\ \bibnamefont
  {Aliev}}, \bibinfo {author} {\bibfnamefont {A.}~\bibnamefont {Ozpineci}},
  \bibinfo {author} {\bibfnamefont {M.}~\bibnamefont {Savci}},\ and\ \bibinfo
  {author} {\bibfnamefont {V.~S.}\ \bibnamefont {Zamiralov}},\ }\bibinfo
  {title} {{Vector meson-baryon strong coupling contants in light cone QCD sum
  rules}},\ \href {https://doi.org/10.1103/PhysRevD.80.016010} {\bibfield
  {journal} {\bibinfo  {journal} {Phys. Rev. D}\ }\textbf {\bibinfo {volume}
  {80}},\ \bibinfo {pages} {016010} (\bibinfo {year} {2009})},\ \Eprint
  {https://arxiv.org/abs/0905.4664} {arXiv:0905.4664 [hep-ph]} \BibitemShut
  {NoStop}%
\bibitem [{\citenamefont {Cheng}\ \emph
  {et~al.}(2025{\natexlab{b}})\citenamefont {Cheng}, \citenamefont {Xu},\ and\
  \citenamefont {Zhong}}]{Cheng:2025oyr}%
  \BibitemOpen
  \bibfield  {author} {\bibinfo {author} {\bibfnamefont {H.-Y.}\ \bibnamefont
  {Cheng}}, \bibinfo {author} {\bibfnamefont {F.}~\bibnamefont {Xu}},\ and\
  \bibinfo {author} {\bibfnamefont {H.}~\bibnamefont {Zhong}},\ }\bibinfo
  {title} {{CP violation in hadronic weak decays of charmed baryons in the
  topological diagrammatic approach}},\ \href
  {https://doi.org/10.1103/n8fk-v7r8} {\bibfield  {journal} {\bibinfo
  {journal} {Phys. Rev. D}\ }\textbf {\bibinfo {volume} {112}},\ \bibinfo
  {pages} {054022} (\bibinfo {year} {2025}{\natexlab{b}})},\ \Eprint
  {https://arxiv.org/abs/2505.07150} {arXiv:2505.07150 [hep-ph]} \BibitemShut
  {NoStop}%
\bibitem [{\citenamefont {Korner}(1971)}]{Korner:1970xq}%
  \BibitemOpen
  \bibfield  {author} {\bibinfo {author} {\bibfnamefont {J.~G.}\ \bibnamefont
  {Korner}},\ }\bibinfo {title} {{Octet behaviour of single-particle matrix
  elements $ \langle B'|H_W|B\rangle $ and $\langle M'|H_W|M\rangle$ using a
  weak current current quark Hamiltonian}},\ \href
  {https://doi.org/10.1016/0550-3213(71)90538-4} {\bibfield  {journal}
  {\bibinfo  {journal} {Nucl. Phys. B}\ }\textbf {\bibinfo {volume} {25}},\
  \bibinfo {pages} {282} (\bibinfo {year} {1971})}\BibitemShut {NoStop}%
\bibitem [{\citenamefont {Pati}\ and\ \citenamefont {Woo}(1971)}]{Pati:1970fg}%
  \BibitemOpen
  \bibfield  {author} {\bibinfo {author} {\bibfnamefont {J.~C.}\ \bibnamefont
  {Pati}}\ and\ \bibinfo {author} {\bibfnamefont {C.~H.}\ \bibnamefont {Woo}},\
  }\bibinfo {title} {{$\Delta I$ = 1/2 rule with fermion quarks}},\ \href
  {https://doi.org/10.1103/PhysRevD.3.2920} {\bibfield  {journal} {\bibinfo
  {journal} {Phys. Rev. D}\ }\textbf {\bibinfo {volume} {3}},\ \bibinfo {pages}
  {2920} (\bibinfo {year} {1971})}\BibitemShut {NoStop}%
\bibitem [{\citenamefont {Aaij}\ \emph
  {et~al.}(2025{\natexlab{b}})\citenamefont {Aaij} \emph
  {et~al.}}]{LHCb:2025hul}%
  \BibitemOpen
  \bibfield  {author} {\bibinfo {author} {\bibfnamefont {R.}~\bibnamefont
  {Aaij}} \emph {et~al.} (\bibinfo {collaboration} {LHCb}),\ }\bibinfo {title}
  {{Amplitude analysis of the ${\ensuremath{\Xi}}_c^+\to
  pK^-{\ensuremath{\pi}}^+$ decay and ${\ensuremath{\Xi}}_c^+$ baryon
  polarization measurement in semileptonic beauty-hadron decays}},\ \href
  {https://doi.org/10.1103/gcft-fgp1} {\bibfield  {journal} {\bibinfo
  {journal} {Phys. Rev. D}\ }\textbf {\bibinfo {volume} {112}},\ \bibinfo
  {pages} {092003} (\bibinfo {year} {2025}{\natexlab{b}})},\ \Eprint
  {https://arxiv.org/abs/2508.00492} {arXiv:2508.00492 [hep-ex]} \BibitemShut
  {NoStop}%
\bibitem [{\citenamefont {Link}\ \emph {et~al.}(2002)\citenamefont {Link} \emph
  {et~al.}}]{FOCUS:2002rvb}%
  \BibitemOpen
  \bibfield  {author} {\bibinfo {author} {\bibfnamefont {J.~M.}\ \bibnamefont
  {Link}} \emph {et~al.} (\bibinfo {collaboration} {FOCUS}),\ }\bibinfo {title}
  {{Measurements of Relative Branching Ratios of $\Lambda^+_c$ Decays into
  States Containing $\Sigma$}},\ \href
  {https://doi.org/10.1016/S0370-2693(02)02103-2} {\bibfield  {journal}
  {\bibinfo  {journal} {Phys. Lett. B}\ }\textbf {\bibinfo {volume} {540}},\
  \bibinfo {pages} {25} (\bibinfo {year} {2002})},\ \Eprint
  {https://arxiv.org/abs/hep-ex/0206013} {arXiv:hep-ex/0206013} \BibitemShut
  {NoStop}%
\bibitem [{\citenamefont {Aaij}\ \emph {et~al.}(2023)\citenamefont {Aaij} \emph
  {et~al.}}]{LHCb:2022sck}%
  \BibitemOpen
  \bibfield  {author} {\bibinfo {author} {\bibfnamefont {R.}~\bibnamefont
  {Aaij}} \emph {et~al.} (\bibinfo {collaboration} {LHCb}),\ }\bibinfo {title}
  {{Amplitude analysis of the ${\ensuremath{\Lambda}}_c^+ \to
  pK^-{\ensuremath{\pi}}^+$ decay and ${\ensuremath{\Lambda}}_c^+$ baryon
  polarization measurement in semileptonic beauty hadron decays}},\ \href
  {https://doi.org/10.1103/PhysRevD.108.012023} {\bibfield  {journal} {\bibinfo
   {journal} {Phys. Rev. D}\ }\textbf {\bibinfo {volume} {108}},\ \bibinfo
  {pages} {012023} (\bibinfo {year} {2023})},\ \Eprint
  {https://arxiv.org/abs/2208.03262} {arXiv:2208.03262 [hep-ex]} \BibitemShut
  {NoStop}%
\bibitem [{\citenamefont {Ablikim}\ \emph
  {et~al.}(2026{\natexlab{b}})\citenamefont {Ablikim} \emph
  {et~al.}}]{BESIII:2026dkj}%
  \BibitemOpen
  \bibfield  {author} {\bibinfo {author} {\bibfnamefont {M.}~\bibnamefont
  {Ablikim}} \emph {et~al.} (\bibinfo {collaboration} {BESIII}),\ }\bibinfo
  {title} {{Amplitude Analysis of Singly Cabibbo-Suppressed Decay
  $\Lambda^{+}_{c}\to p K^{+} K^{-}$}},\ \href@noop {} {\  (\bibinfo {year}
  {2026}{\natexlab{b}})},\ \Eprint {https://arxiv.org/abs/2603.08469}
  {arXiv:2603.08469 [hep-ex]} \BibitemShut {NoStop}%
\bibitem [{\citenamefont {Ablikim}\ \emph {et~al.}(2025)\citenamefont {Ablikim}
  \emph {et~al.}}]{BESIII:2024xny}%
  \BibitemOpen
  \bibfield  {author} {\bibinfo {author} {\bibfnamefont {M.}~\bibnamefont
  {Ablikim}} \emph {et~al.} (\bibinfo {collaboration} {BESIII}),\ }\bibinfo
  {title} {{Measurement of the branching fractions of the decays
  ${\ensuremath{\Lambda}}_c^+{\rightarrow}{\ensuremath{\Lambda}}K_S^0 K^+,
  {\ensuremath{\Lambda}}_c^+{\rightarrow}{\ensuremath{\Lambda}}K_S^0{\ensuremath{\pi}}^+$,
  and ${\ensuremath{\Lambda}}_c^+{\rightarrow}{\ensuremath{\Lambda}}K^{*+}$}},\
  \href {https://doi.org/10.1103/PhysRevD.111.012014} {\bibfield  {journal}
  {\bibinfo  {journal} {Phys. Rev. D}\ }\textbf {\bibinfo {volume} {111}},\
  \bibinfo {pages} {012014} (\bibinfo {year} {2025})},\ \Eprint
  {https://arxiv.org/abs/2410.16912} {arXiv:2410.16912 [hep-ex]} \BibitemShut
  {NoStop}%
\bibitem [{\citenamefont {Aaij}\ \emph {et~al.}(2024)\citenamefont {Aaij} \emph
  {et~al.}}]{LHCb:2024hju}%
  \BibitemOpen
  \bibfield  {author} {\bibinfo {author} {\bibfnamefont {R.}~\bibnamefont
  {Aaij}} \emph {et~al.} (\bibinfo {collaboration} {LHCb}),\ }\bibinfo {title}
  {{Search for the rare decay of charmed baryon ${\ensuremath{\Lambda}}_c^+$
  into the $p{\ensuremath{\mu}}^+ {\ensuremath{\mu}}^-$ final state}},\ \href
  {https://doi.org/10.1103/PhysRevD.110.052007} {\bibfield  {journal} {\bibinfo
   {journal} {Phys. Rev. D}\ }\textbf {\bibinfo {volume} {110}},\ \bibinfo
  {pages} {052007} (\bibinfo {year} {2024})},\ \Eprint
  {https://arxiv.org/abs/2407.11474} {arXiv:2407.11474 [hep-ex]} \BibitemShut
  {NoStop}%
\bibitem [{\citenamefont {Link}\ \emph {et~al.}(2001)\citenamefont {Link} \emph
  {et~al.}}]{FOCUS:2001ovr}%
  \BibitemOpen
  \bibfield  {author} {\bibinfo {author} {\bibfnamefont {J.~M.}\ \bibnamefont
  {Link}} \emph {et~al.} (\bibinfo {collaboration} {FOCUS}),\ }\bibinfo {title}
  {{Measurement of the Relative Branching Ratio BR($\Xi^+_c \to p^+ K^- \pi^+$)
  / BR($\Xi^+_c \to \Xi^- \pi^+ \pi^+$)}},\ \href
  {https://doi.org/10.1016/S0370-2693(01)00590-1} {\bibfield  {journal}
  {\bibinfo  {journal} {Phys. Lett. B}\ }\textbf {\bibinfo {volume} {512}},\
  \bibinfo {pages} {277} (\bibinfo {year} {2001})},\ \Eprint
  {https://arxiv.org/abs/hep-ex/0102040} {arXiv:hep-ex/0102040} \BibitemShut
  {NoStop}%
\end{thebibliography}%
\end{document}